\documentclass{article}

\usepackage{lmodern}
\usepackage{arxiv}

\usepackage[utf8]{inputenc} % allow utf-8 input
\usepackage[T1]{fontenc}    % use 8-bit T1 fonts
\usepackage{hyperref}       % hyperlinks
\usepackage{url}            % simple URL typesetting

\usepackage{nicefrac}       % compact symbols for 1/2, etc.
\usepackage{microtype}      % microtypography

\usepackage{graphicx}
\usepackage[round]{natbib}
\usepackage{setspace}
\usepackage{amsfonts}
\usepackage{amsmath}
\usepackage{amssymb}
\usepackage{mathrsfs}
\usepackage{nameref}
\usepackage{appendix}
\usepackage{tabularx}
\usepackage{siunitx}
\usepackage{multirow}
\usepackage{bm}
\usepackage{threeparttable}
\usepackage{pdflscape}
\usepackage{rotating}
\usepackage{booktabs}

\usepackage{caption}
\usepackage{subcaption}
\usepackage{mathtools}
\mathtoolsset{showonlyrefs=true}

\usepackage{algorithm}

\usepackage[export]{adjustbox}

\renewcommand{\headeright}{}
\renewcommand{\undertitle}{}

\date{December 30, 2022}

\begin{document}

\title{Dynamic Term Structure Models with Nonlinearities using Gaussian Processes}

\author{Tomasz Dubiel-Teleszynski \\
	Department of Statistics, London School of Economics\\
	Houghton Street, London, WC2A 2AE, United Kingdom \\
	\texttt{t.dubiel-teleszynski1@lse.ac.uk} \\
	%% examples of more authors
	\And
	Konstantinos Kalogeropoulos \\
	Department of Statistics, London School of Economics\\
	Houghton Street, London, WC2A 2AE, United Kingdom \\
	\texttt{k.kalogeropoulos@lse.ac.uk} \\
	\And
	Nikolaos Karouzakis \\
	ALBA Graduate Business School, The American College of Greece\\
	6-8 Xenias Str, 115 28 Athens, Greece \\
	\& University of Sussex Business School, Brighton, United Kingdom \\
	\texttt{nkarouzakis@alba.acg.edu} \\
}

\maketitle

\begin{abstract}
The importance of unspanned macroeconomic variables for Dynamic Term Structure Models has been intensively discussed in the literature. To our best knowledge the earlier studies considered only linear interactions between the economy and the real-world dynamics of interest rates in DTSMs. We propose a generalized modelling setup for Gaussian DTSMs which allows for unspanned nonlinear associations between the two and we exploit it in forecasting. Specifically, we construct a custom sequential Monte Carlo estimation and forecasting scheme %based on \cite{Chopin02,Chopin04}
where we introduce Gaussian Process priors to model nonlinearities. Sequential scheme we propose can also be used with dynamic portfolio optimization to assess the potential of generated economic value to investors. The methodology is presented using US Treasury data and selected macroeconomic indices. Namely, we look at core inflation and real economic activity. %as in \cite{Cieslak15} and \cite{Joslin14}, respectively.
We contrast the results obtained from the nonlinear model with those stemming from an application of a linear model. %which is conceptually closest to \cite{Joslin14}.
Unlike for real economic activity, in case of core inflation we find that, compared to linear models, application of nonlinear models leads to statistically significant gains in economic value across considered maturities.

\end{abstract}

\newpage

% \listofalgorithms
% \newpage
% \listoffigures
% \newpage
% \listoftables
% \newpage

\section{Introduction}
\label{Intro}

\subsection{Term Structure and Macroeconomic Information}% and Nonlinearities}

A fundamental and economically important argument in macro-finance literature suggests that the current yield curve spans all relevant information for forecasting future yields, returns and bond risk premia. Although the argument is implied by most macro-finance models, recent evidence raises important questions on its validity. In particular, several empirical studies (see, \cite{Cooper09}, \cite{Ludvigson09}, \cite{Duffee11}, \cite{Joslin14}, \cite{Cieslak15}, \cite{Gargano19}, \cite{Bianchi20}) provide support for unspanned macroeconomic risks\footnote{e.g.: the output gap of \cite{Cooper09}, the ‘real’ and ‘inflation’ factors of \cite{Ludvigson09}, the measures of economic activity and inflation of \cite{Joslin14} and the long-run inflation expectation of \cite{Cieslak15}, among others.} possessing considerable predictive power above and beyond the yield curve. The vast majority of the extant literature, however, assumes that the relationship between those unspanned macroeconomic factors and the cross section of yields is mostly, if not purely, linear in nature. 

\cite{Duffee11} is the first to comment on the possibility of a nonlinear (albeit, economically weak) relationship between bond risk premia and macroeconomic activity. Only recently, \cite{Bianchi20} provide evidence on the importance of accounting for nonlinearities in order to detect further information important for forecasting bond risk premia. Due to the rapid increase in the use of machine learning methods, the presence of nonlinearities, relevant for predicting excess returns, has recently attracted attention in the literature. However, we are unaware of any prior research that explores the possibility of an asymmetric/nonlinear relationship between unspanned macroeconomic risks and return predictability in the US bond market within an arbitrage-free pricing model. In this paper, we seek to investigate this relationship, if any, further. With this in mind, we propose a novel class of arbitrage-free Dynamic Term Structure Models (DTSM) that embed nonlinear unspanned macroeconomic risks using Gaussian Processes \citep{Bishop2006,RasmussenWilliams2006}.

%In this paper, we seek to investigate further the possibility of an asymmetric or nonlinear relation between unspanned macroeconomic risks and return predictability in the US bond market. 

This paper contributes to the literature that studies the linkages between bond risk premia and the macroeconomy. Our first contribution is methodological. Following \cite{Joslin14}, we embed unspanned information coming from (directly observable) macroeconomic risks within a DTSM. The point where our approach deviates from prior studies is that this information is assumed exogenous and is entering the model in a nonlinear fashion without making specific assumption on the manner it affects the yield curve. To do so, we introduce a novel methodological framework that utilises Bayesian non-parametrics within a DTSM, thus staying agnostic about the functional form between macroeconomic activity and bond risk premia. Our setup, also allows us to handle Gaussian Processes sequentially and effectively perform tasks such as sequential parameter estimation and forecasting. Drawing from the work of \cite{Chopin02} and \cite{DelMoral06}, we allow for a sequential Bayesian treatment of exogeneous macroeconomic information in a nonlinear manner.

Our second contribution is empirical. From a fundamental and economic perspective, the proposed framework seeks to enhance our understanding behind the determinants of bond risk premia. In particular, whether movements in excess bond returns bear any relation to the macroeconomy. Are there important driving forces of bond return predictability that are nonlinear and hard, if not impossible, to summarize utilising information coming solely from linear transformations of the data? Furthermore, we seek to investigate deeper whether such nonlinearities, if any, allow investors to exploit the evident statistical predictability of the resulting models and generate economically significant portfolio benefits to bond investors, out-of-sample. 

\subsection{Economic Benefits from Nonlinearities}

Our results reveal a direct line linking the nonlinear component of the unspanned measures of macroeconomic activity to model performance, especially when it comes to generating additional economic value. In order to scrutinize nonlinear models in a meaningful way, as benchmark we apply a version of the macro-finance DTSM proposed by \cite{Joslin14}, which incorporates unspanned macroeconomic variables in a linear manner. However, we proceed without modeling their underlying dynamics, in order to remain closest possible to the nonlinear setup we propose.

First, we find that nonlinear models provide a competitive edge over linear models in cases where \textit{ex-ante} nonlinear functions of macroeconomic variables preserve their nonlinear nature \textit{ex-post}, in the part which is hidden from the yield curve. In this paper, these are specifically the prices, and not the economic activity, which we also consider. We demonstrate that by resorting to risk premium factor decomposition from \cite{Duffee11}.

Second, results in this paper also indicate that for some models, where macroeconomic risk directly affects the level of the yield curve, a nonlinear formulation, like ours which relies on Gaussian Processes, does not necessarily mean outstanding model performance in terms of predictability and economic value gains, however it lets avoid substantial deterioration on that front. In particular, this is the case for macroeconomic variables which in abnormal times, such as the $2008$-$2009$ financial crisis or the more recent COVID-$19$ recession, are subject to outliers, like in this paper the economic activity.

%\subsection{Bond Predictability and Economic Benefits}

\subsection{Outline}

The remainder of this paper is organized as follows. Section \ref{GPsection} is a brief introduction to Gaussian Processes. Section \ref{Model} describes the proposed modelling framework. Section \ref{sec:Sequential} presents the procedure for sequential learning with Gaussian Processes and forecasting, along with the framework for assessing the predictive and economic performance of the resulting models. Section \ref{sec:Data} discusses the data and the sample period used and presents the family of models considered in this paper. Section \ref{sec:application} discusses the results both in terms of predictive performance and economic value, including the associated explanatory power where applicable, as well as reveals the links between these results and the hidden nonlinearities. Finally, Section \ref{sec:Conclude} concludes the paper by providing some relevant discussion.

\section{Gaussian Processes}
\label{GPsection}

In Bayesian non-parametrics a Gaussian Process prior is used to estimate unknown function in supervised learning setting. One does not assume a specific function between some $y$ and $x$ and instead performs Bayesian inference on the function. Univariate exposition below follows from \cite{Bishop2006}, \cite{RasmussenWilliams2006} and \cite{Murphy2012}.

\subsection{Gaussian Process Theory}

Let $\mathscr{X}$ be a set and $\mathscr{V}$ be a set of functions over $\mathscr{X}$, for example smooth functions. We observe $(x_1,y_1)$ ...  $(x_t,y_t)$ ... $(x_T,y_T)$, where $x_t \in \mathscr{X}$, $y_t \in \mathscr{R}$, $t=1,...,T$, satisfying

\begin{equation}
y_t = v(x_t) + \epsilon_t
\end{equation}

\noindent where $v \in \mathscr{V}$ and $\epsilon=(\epsilon_1,...,\epsilon_T)$ are independent of $x=(x_1,...,x_T)$. Errors $\epsilon$ have density $\pi(\mathbf{\epsilon})$ and usually it is $N(0_T,\; \sigma_{\epsilon}^2 I_T$), $0_T$ being $(T\times 1)$ vector of zeros, what further implies that

\begin{equation}
y = (y_1,...,y_T)|v,x,\sigma_{\mathbf{\epsilon}}^2 \sim N(v,\; \sigma_{\epsilon}^2 I_T)
\end{equation}

\noindent where $v=[v(x_1),...,v(x_T)]'$. In a Bayes regression we assign prior $\pi(v)$ to $v$ and compute the posterior

\begin{equation}
\pi(v|y) = \frac{\pi(v)f(\mathbf{\epsilon}|v)}{\int_{\mathscr{V}}\pi(v)f(\mathbf{\epsilon}|v)dv}
\end{equation}

\noindent where $f(\mathbf{\epsilon}|v)$ is the likelihood. Then $v$ can be estimated by its posterior mean.

\medskip
\noindent $\bf Definition$. Let $\mathscr{X}$ be a set. A random function $v:\mathscr{X}\to\mathscr{R}$ is called a Gaussian Process ($GP$) if for any $x_1,...,x_T\in\mathscr{X}$, $v=[v(x_1),...,v(x_T]'$ has a multivariate normal distribution.
\medskip

Gaussian Process is characterized by the mean $v_0(\cdot)$ and the covariance kernel $K(\cdot,\cdot)$. The latter is a positive definite function $K:\mathscr{X}\times\mathscr{X}\to\mathscr{R}$. If $v$ is a $GP$ with mean $v_0$ and covariance kernel $K$ then

\begin{equation}
v=[v(x_1),...,v(x_T]' \sim N(v_0,\; K)
\end{equation}

\noindent where $v_0=[v_0(x_1),...,v_0(x_T]$ and $K$ is $(T \times T)$ matrix with elements $K(s,t)$ such that $s,t=1,...,T$.
There are numerous possibilities when it comes to covariance kernels and their choice depends on the application. For instance, they can be linear kernels

\begin{equation}
K(x,x') = \left\langle x,x' \right\rangle
\end{equation}

\noindent or Gaussian kernels

\begin{equation}
K(x,x') = e^{-\frac{\left\|x-x'\right\|^2}{2\sigma^2}}
\end{equation}

\noindent among others. For details about these and other kernels, and how they can be combined, see \cite{RasmussenWilliams2006}.

\subsection{Gaussian Process Regression}

For $t=1,...,T$ assume that

\begin{equation}
y_t = v(x_t) + \epsilon_t
\end{equation}

\noindent with

$$
\begin{array}{c}
\pi(v) = GP(v_0,\; K) \\
\epsilon_t\sim N(0,\; \sigma_{\epsilon}^2) 
\end{array}
$$

\noindent where $v$ and $(\epsilon_1,...,\epsilon_T)$ are independent and for simplicity one can assume that $v_0=0$, what is also a common assumption made in applications.

Marginal distribution of $y$ is thus multivariate normal with means $E(y_t)=0$ and covariances $Cov(y_s,y_t)=K(x_s,x_t) + \sigma_{\epsilon}^2 I(s=t)$, with $s,t=1,...,T$. In matrix notation, where $K_T$ denotes the $(T\times T)$ matrix containing all the $(x_s,x_t)$ pairs, we get
\begin{equation}
y \sim N(0_{T\times1},\; K_T + \sigma_{\epsilon}^2 I_T)
\end{equation}
Since for each $x_t \in \mathscr{X}$ and $y_t$ we have that $Cov(v(x_s),y_t)=K(x_s,x_t)$, the joint distribution of $v$ and $y$ is

\begin{equation}
\left(
    \begin{array}{c}
    v \\
    y
    \end{array}
\right) \sim N \left[
    \left(\begin{array}{c} 
    0_{T\times1} \\
    0_{T\times1}
    \end{array}\right),\; \left(
    \begin{matrix}
    K_T & K_T \\
    K_T & K_T + \sigma_{\epsilon}^2 I_T
    \end{matrix}
    \right)
\right]
\end{equation}

\noindent where $v=[v(x_1),...,v(x_T)]'$ and $0_{T\times1}$ is a $(T\times 1)$ vector of zeros.

Under such assumptions, by the multivariate normal regression lemma \citep{DurbinKoopman2012}, the distribution of $v|y$ is a Gaussian with mean $\mu_v$ and covariance kernel $\Sigma_v$, which are following

$$
\begin{array}{l}
\mu_v = K_T[K_T+\sigma_{\epsilon}^2 I_T]^{-1}y \\
\Sigma_v = K_T-K_T[K_T+\sigma_{\epsilon}^2 I_T]^{-1}K_T
\end{array}
$$

To project $y_{T+1}$ based on $x_{T+1}$ we need to find the joint distribution of $y$ and $y_{T+1}$, given $x$ and $x_{T+1}$. To that end, we note that $Cov(y_t,y_{T+1})=K(x_t,x_{T+1})$ for $t=1,...,T$ and we let 
$$
k_{T+1}=[K(x_1,x_{T+1}) ... K(x_T,x_{T+1})]'
$$
to arrive at

\begin{equation}
\left(
    \begin{array}{c}
    y_{T+1} \\
    y
    \end{array}
\right) \sim N \left[
    \left(\begin{array}{c}
    0 \\
    0_{T\times1}
    \end{array}\right),\; \left(
    \begin{matrix}
    K(x_{T+1},x_{T+1})+\sigma_{\epsilon}^2 & k_{T+1}' \\
    k_{T+1} & K_T + \sigma_{\epsilon}^2 I_T
    \end{matrix}
    \right)
\right]
\end{equation}

\noindent Then, by the multivariate normal regression lemma, $y_{T+1}|y$ is a Gaussian with mean $\mu_{T+1}$ and variance $\Sigma_{T+1}$ such that

$$
\begin{array}{l}
\mu_{T+1} = k_{T+1}'[K_T+\sigma_{\epsilon}^2 I_T]^{-1}y \\
\Sigma_{T+1} = K(x_{T+1},x_{T+1})+\sigma_{\epsilon}^2-k_{T+1}'[K_T+\sigma_{\epsilon}^2 I_T]^{-1}k_{T+1}
\end{array}
$$

Posterior distribution of $v$ still depends on unknown vector of parameters $\mathbf{\theta}$ consisting of $\sigma_{\epsilon}$ and any hyper-parameters of $K$, for example $\sigma$ for squared exponential kernel. In principle, two methods can be used to estimate $\mathbf{\theta}$. We can maximize the marginal likelihood $\pi(y|\theta)$ which is a multivariate normal density with mean $\mu_y=0_{T\times 1}$ and covariance matrix $\Sigma_y=K_{T}+\sigma_{\epsilon}^2 I_T$. Doing so is denoted empirical Bayes. We can also put a prior on the hyper-parameters and estimate them by their posterior means. Doing this is known as hierarchical Bayes \citep{Bishop2006,Murphy2012}. In what follows, we pursue a pragmatic combination of the two approaches.

\section{Dynamic Term Structure Model, Likelihood, and Nonlinear Macros}
\label{Model}

%Section \ref{subsec:basecase} below contains the standard DTSM, which is also the starting point in \cite{DTKK2021b}. It only appears here to create a self contained paper, thus the readers are encouraged to skip directly to Section \ref{subsec:UnspannedMacros}, if they are already familiar with it.

\subsection{Base Case}
\label{subsec:basecase}

To start with, consider the no-arbitrage class of discrete-time Affine Term Structure Models (ATSM) (see, \cite{Ang03} and \cite{Cochrane05}) and in particular the Gaussian case. Under the physical probability measure $\mathbb{P}$, let us consider first that a $(N \times 1)$ vector of state variables $X_t$, $t=0,1,...,T$, evolves according to a first-order Gaussian Vector Autoregressive (VAR) process 
\begin{equation}\label{VAR_P}
X_{t}= \mu^{\mathbb{P}} + \Phi^{\mathbb{P}} X_{t-1} + \Sigma \varepsilon_{t}
\end{equation}
where $\varepsilon_{t} \sim N(0,\; I_{N})$, $\Sigma$ is a $(N \times N)$ lower triangular matrix, $\mu^{\mathbb{P}}$ is a $(N \times 1)$ vector and $\Phi^{\mathbb{P}}$ is a $(N \times N)$ matrix. Under this initial framework, the one period risk-free interest rate $r_t$\footnote{Working with monthly data implies that $r_t$ is the 1-month yield.} is assumed to be an affine function of the state variables
\begin{equation}\label{short rate}
r_{t}=\delta_{0} + \delta_{1}' X_{t}
\end{equation}
where $\delta_{0}$ is a scalar and $\delta_{1}$ is a $(N \times 1)$ vector. Absence of arbitrage implies the existence of a pricing kernel $\mathcal{M}_{t+1}$ specified as
\begin{equation}
\label{pricingkernel}
\mathcal{M}_{t+1} =  \exp( -r_{t}-\frac{1}{2} \lambda_{t}' \lambda_{t} - \lambda_{t}' \varepsilon_{t+1})
\end{equation}
with $\lambda_{t}$ being the time-varying market prices of risk, which is also assumed to be affine in the state vector $X_t$
%\footnote{This is the 'essentially-affine' specification introduced in \cite{Duffee02}. Existing studies have proposed alternative specifications for the market price of risk, such as the 'completely-affine' model of \cite{Dai00}, the 'semi-affine' model of \cite{Duarte04} and the 'extended-affine' model of \cite{Cheridito07}. See, \cite{Feldhutter16} for a useful comparison of the models.},
\begin{equation}\label{price of risk}
\lambda_{t} = \Sigma^{-1}\left(\lambda_{0} + \lambda_{1} X_{t}\right)
\end{equation}
where $\lambda_{0}$ is a $(N \times 1)$ vector and $\lambda_{1}$ is a $(N \times N)$ matrix.
If we assume that the pricing kernel $\mathcal{M}_{t+1}$ prices all bonds in the economy and we let $P_{t}^{n}$ denote the time-t price of an n-period zero-coupon bond, then the price of the bond is computed from $P_{t}^{n}=E_{t} \left ( \mathcal{M}_{t+1}P_{t+1}^{n-1} \right )$ where $E_{t}(\cdot)$ denotes expectation given the information available up to time $t$. As such, it follows that bond prices are exponentially affine functions of the state vector (see, \cite{Duffie96})
% \begin{equation}
% P_{t}^{n+1}=E_{t}^{\mathbb{Q}} \left [ M_{t+1}P_{t+1}^{n} \right ]
% \end{equation}
\begin{equation}\label{bond price}
P_{t}^{n} = \exp(A_{n} + B_{n}' X_{t})
\end{equation}
with loadings, $A_{n}$ being a scalar and $B_{n}$ a $(N \times 1)$ vector, satisfying the following recursions
\begin{align}\label{recursions} 
A_{n+1} &= A_{n} + B_{n}' ( \mu - \lambda_{0} ) + \frac{1}{2} B_{n}' \Sigma \Sigma' B_{n} - \delta_{0} \\ 
B_{n+1} &=  ( \Phi - \lambda_{1} )'B_{n} - \delta_{1}
\end{align}
with $A_{1} = -\delta_{0}$ and $B_{1} = -\delta_{1}$. The $\mathbb{Q}$ dynamics of the state vector are given by
\begin{equation}\label{VAR_Q}
X_{t}= \mu^{\mathbb{Q}} + \Phi^{\mathbb{Q}} X_{t-1} + \Sigma \varepsilon_{t}^{\mathbb{Q}}
\end{equation}
where $\mu^{\mathbb{Q}} = \mu - \lambda_{0}$, $\Phi^{\mathbb{Q}} = \Phi - \lambda_{1}$ and $\varepsilon_{t}^{\mathbb{Q}} \sim N(0,\; I_{N})$.
The continuously compounded n-period yield $y_t^n$ is also an affine function of the state vector
\begin{equation}\label{oldyield}
y_{t}^{n} = - \frac{\log P_{t}^{n}}{n} = A_{n,X}+B_{n,X}'X_{t}
\end{equation}
where the loading scalar $A_{n,X}$ and the loading $(N \times 1)$ vector $B_{n,X}$ are calculated using the above recursions as $A_{n,X}=-A_{n}/n$ and $B_{n,X} = -B_{n}/n$.

Next, we follow the canonical setup of \cite{Joslin11} and rotate the vector of unobserved state variables $X_t$ such that they are linear combinations of the observed yields. In particular, we rotate $X_t$ to the first $N$ principal components (PCs) of observed yields as
\begin{equation}\label{PCs}
\mathcal{P}_{t} = W y_t = WA_{X} + WB_{X}X_{t}
\end{equation}
where $W$ denotes a $(N \times J)$ matrix containing first $N$ PCs' loadings, $y_t$ is a $(J \times 1)$ vector of continuously compounded yields. The $(J\times 1)$ vector $A_X$ and the $(J\times N)$ matrix $B_X$ contain model-implied loadings of yields on risk factors, specifically $A_{n,X}$ are elements of vector $A_X$ and $B_{n,X}$ are transposed rows of matrix $B_X$. This allows us to re-write the yield equation (\ref{oldyield}) in vector form
\begin{equation} \label{oldyieldvec}
y_{t} = A_{X}+B_{X}X_{t}
\end{equation}
and then as a function of the observable risk factors $\mathcal{P}_{t}$
\begin{equation}\label{newyield}
y_{t} = A_{\mathcal{P}}+B_{\mathcal{P}} \mathcal{P}_{t}
\end{equation}
where the loadings $A_{\mathcal{P}}$ and $B_{\mathcal{P}}$, which are derived by applying transformation in \eqref{PCs} to \eqref{oldyieldvec}, are given below\footnote{According to \cite{Duffee11}, outside of knife-edge cases the matrix $(W B_{X})$ is invertible, and as such $\mathcal{P}_{t}$ contains the same information as $X_t$.}
\begin{align}
A_{\mathcal{P}} &= A_{X} - B_{X} ( W B_{X} )^{-1} W A_{X} \label{rotation2} \\ 
B_{\mathcal{P}} &=  B_{X} ( W B_{X} )^{-1}\label{rotation2_2} 
\end{align}
Furthermore, the risk-neutral dynamics of $\mathcal{P}_{t}$ are given as
\begin{equation}\label{VAR_P_Q}
\mathcal{P}_{t}= \mu_{\mathcal{P}}^{\mathbb{Q}} + \Phi_{\mathcal{P}}^{\mathbb{Q}} \mathcal{P}_{t-1} + \Sigma_{\mathcal{P}} \varepsilon_{t}^{\mathbb{Q}}
\end{equation}
where the risk-neutral measure parameters $\mu_{\mathcal{P}}^{\mathbb{Q}}$, $\Phi_{\mathcal{P}}^{\mathbb{Q}}$ and $\Sigma_{\mathcal{P}}$ are derived similarly from \eqref{VAR_Q}, what leads to
\begin{align}\label{rotation1} 
\mu_{\mathcal{P}}^{\mathbb{Q}} &= W B_{X}\mu^{\mathbb{Q}} +(I_N - \Phi_{\mathcal{P}}^{\mathbb{Q}})W A_{X} \\ 
\Phi_{\mathcal{P}}^{\mathbb{Q}} &= W B_{X} \Phi^{\mathbb{Q}} (W B_{X})^{-1}\\
\Sigma_{\mathcal{P}} &= W B_{X} \Sigma 
\end{align}
Finally, given the new observable state vector $\mathcal{P}_{t}$, the one period short rate $r_t$ is also an affine function of $\mathcal{P}_{t}$ given as
\begin{equation}\label{rateP}
r_{t}=\delta_{0\mathcal{P}} + \delta_{1\mathcal{P}}' \mathcal{P}_{t}
\end{equation}
with 
\begin{align}\label{rotation3} 
\delta_{0\mathcal{P}} &= \delta_{0} - \delta_{1}'(W B_{X})^{-1}W A_{X} \\ 
\delta_{1\mathcal{P}}' &= (W B_{X})^{-1} \delta_{1}
\end{align}
and the market price of risk specification becomes accordingly
\begin{equation}\label{rotation4}
\lambda_{t} = \Sigma_\mathcal{P}^{-1}\left(\lambda_{0\mathcal{P}} + \lambda_{1\mathcal{P}} \mathcal{P}_{t}\right)
\end{equation}
where
\begin{align} 
\lambda_{0\mathcal{P}} &= W B_{X}\lambda_{0} - W B_{X}\lambda_{1}(W B_{X})^{-1}W A_{X}\label{lam0p} \\ 
\lambda_{1\mathcal{P}} &= W B_{X}\lambda_{1} (W B_{X})^{-1} \label{lam1p}
\end{align}
We note that $\mathbb{P}$ dynamics of $\mathcal{P}_t$ are of equivalent form to \eqref{VAR_P_Q} with
\begin{align}
\mu_{\mathcal{P}}^{\mathbb{P}}&=\mu_{\mathcal{P}}^{\mathbb{Q}}+\lambda_{0\mathcal{P}}\label{PPconst} \\ \Phi_{\mathcal{P}}^{\mathbb{P}}&=\Phi_{\mathcal{P}}^{\mathbb{Q}}+\lambda_{1\mathcal{P}}\label{PPfeed}
\end{align}
where $\lambda_{0\mathcal{P}}$ is a $(N \times 1)$ vector and $\lambda_{1\mathcal{P}}$ is a $(N \times N)$ matrix specified in \eqref{lam0p} and \eqref{lam1p} respectively, reflecting the market price of risk in $\mathcal{P}_{t}$ terms, as well as with $\varepsilon_{t}$ instead of $\varepsilon_{t}^{\mathbb{Q}}$.

Notice also that in \eqref{oldyieldvec} and \eqref{newyield}, yields are assumed to be observed without any measurement error. Nevertheless, an $N$-dimensional observable state vector cannot perfectly price $J>N$ yields, and as such, we further assume that the $(J-N)$ bond yields used in the estimation are observed with independent $N(0,\; \sigma_e^2)$ measurement errors. An equivalent way to formulate this is to write
\begin{equation}
\label{Qmodel}
y_{t}=A_{\mathcal{P}}+B_{\mathcal{P}} \mathcal{P}_{t} + e_t
\end{equation}
and to consider the dimension of $e_t$ as effectively being $(J-N)\times 1$. Letting $W_{\bot}$ denote a basis of the null space of $W$, the measurement error assumption can be also expressed as
\begin{equation}\label{Wbotet}
W_{\bot}e_t \sim N(0,\;\sigma_e^2I_{J-N})
\end{equation}
where $(W_{\bot}e_t)$ is a $(J-N)\times 1$ vector \citep{Bauer18}.

In our setting, we also follow one of the identification schemes proposed in \cite{Joslin11} (see, Proposition 1), where the short rate is the sum of the state variables, given as $r_t = i X_t$ with $i$ being a vector of ones, and the parameters $\mu^{\mathbb{Q}}$ and $\Phi^{\mathbb{Q}}$ of the state vector's $\mathbb{Q}$-dynamics are given as $\mu^{\mathbb{Q}} = [k_{\infty}^{\mathbb{Q}}, 0, 0]$ and $\Phi^{\mathbb{Q}} = diag(g^{\mathbb{Q}})$, where $g^{\mathbb{Q}}$ denotes a $(N \times 1)$ vector containing the real and distinct eigenvalues of $\Phi^{\mathbb{Q}}$ \footnote{Alternative specifications for the eigenvalues are considered in \cite{Joslin11}, however real eigenvalues are found to be empirically adequate.}.

\subsection{Incorporating Unspanned Nonlinear Macros}
\label{subsec:UnspannedMacros}

We further consider an extension of the model presented in the previous section. Our approach resembles the framework of \cite{Joslin14} in that the  model is factorised into a `spanned' component, i.e. risk factors which can be retrieved by the information provided in historical yield curve data, as well as an `unspanned' component that could include background factors such as macroeconomic variables. It is assumed that the latter is not determined by the yield curve, yet it remains highly relevant for the inference and, more importantly, prediction purposes. The points where our approach differs from \cite{Joslin14} are as follows. First, the unspanned components are regarded as unknown, possibly nonlinear functions, which are to be estimated, of observable macroeconomic variables, rather than just macros entering the model in a linear manner. Having said that, the latter case is considered too, as benchmark in model comparisons, and is henceforth referred to as the linear model. Second, the unspanned components are not assumed endogenous but exogenous.

We thus introduce the following nonlinear model for the $\mathbb{P}$-dynamics of the state vector
\begin{equation}
\label{PmodelExog}
\mathcal{P}_t = \mu_{\mathcal{P}}^{\mathbb{P}} + \Phi_{\mathcal{P}}^{\mathbb{P}}\mathcal{P}_{t-1} + v( M_{t-1}) +\Sigma_{\mathcal{P}} \varepsilon_{t}
\end{equation}
where $\mu_{\mathcal{P}}^{\mathbb{P}}$ and $\Phi_{\mathcal{P}}^{\mathbb{P}}$ are defined in \eqref{PPconst} and \eqref{PPfeed} respectively. The matrix $\Sigma_{\mathcal{P}}$ represents the Cholesky decomposition (i.e. lower triangular with positive diagonal elements) of the covariance matrix $\Sigma_{\mathcal{P}}\Sigma_{\mathcal{P}}^{\prime}$, whereas $\varepsilon_{t}$ is a vector of $N$ error terms that are assumed to be normally distributed with zero mean and $I_N$, the identity matrix of dimension $N$, as the covariance matrix. The term $v(M_{t-1})$ is of dimension $(N\times 1)$ and reflects the impact of $R$ (lagged) macros in $M_{t-1}$, a $(R\times 1)$ vector, on $\mathcal{P}_t$, with the function $v(\cdot)$ being such that $v:\mathbb{R}^{R}\rightarrow \mathbb{R}^{N}$. More details are provided in the next section.

The model is completed by setting the $\mathbb{Q}$-dynamics for $\mathcal{P}_t$ as in \eqref{VAR_P_Q}. To ensure that macros are not spanned by $\mathcal{P}_t$ we set $r_t$ as in \eqref{rateP} and the pricing kernel as in \eqref{pricingkernel}, however 
% \begin{equation*}
% \mathcal{M}_{t+1} =  \exp( -r_{t}-\frac{1}{2} \lambda_{t}' \lambda_{t} - \lambda_{t}' \varepsilon_{t+1})
% \end{equation*}
the time-varying market prices of risk $\lambda_{t}$ are now given by 
\begin{align}
\lambda_{t}&=\Sigma_{\mathcal{P}}^{-1}\left[\mu_{\mathcal{P}}^{\mathbb{P}}-\mu_{\mathcal{P}}^{\mathbb{Q}}+(\Phi_{\mathcal{P}}^{\mathbb{P}}\mathcal{} -\Phi_{\mathcal{P}}^{\mathbb{Q}})\mathcal{P}_t+v(M_{t-1})\right] \\
&=\Sigma_{\mathcal{P}}^{-1}\left[\lambda_{0\mathcal{P}}+\lambda_{1\mathcal{P}}\mathcal{P}_t+v(M_{t-1})\right] \label{lambdat}
\end{align}
%
%Note that since macros are not part of the pricing procedure their risk-neutral dynamics is not identified either. Further details on the linear model are in Appendix \ref{appendix:LinearModel}.

%Final model specification assumes appropriate choice of restrictions on $\lambda_{0\mathcal{P}}$ and $\lambda_{1\mathcal{P}}$ in $\lambda_t$, as discussed in Section \ref{subsec:likelihoodGaussian}, and Gaussian Process outputs in $v(M_{t-1})$, as explained in the next section, where the latter depend on observed macroeconomic variables described in Section \ref{sec:Data}.

\subsection{Gaussian Process Mean Ornstein-Uhlenbeck Model}
\label{subsec:GPOUModel}
In what follows, we define in more detail the function $v(\cdot)$ in \eqref{PmodelExog}, and the underlying hyper-parameters. To that end, we follow a Bayesian non-parametric approach using a Gaussian Process with multiple outputs. We focus on a special case of \eqref{PmodelExog} with $N=3$ and $R=1$, noting that the first three extracted PCs are typically sufficient to capture most of the variation in the yield curve and often correspond to its level, slope, and curvature, respectively \citep{Litterman91}.

For interpretation reasons, we work with models containing $R=1$ macro, and explore their submodels. We believe that it is the natural step to gain the understanding, as to how these models behave, before moving to more complex specifications. Nevertheless, the computational scheme we develop in this paper is capable of handling $R>1$ macros in a single model.

%The parsimonious representation of \cite{Joslin11} is adopted by defining $\mu_{\mathcal{P}}^{\mathbb{P}}$ and $\Phi_{\mathcal{P}}^{\mathbb{P}}$ based on their $\mathbb{Q}$ counterparts. It is further completed with an appropriate choice of restrictions on $\lambda_t$, in particular on $\lambda^{\mathcal{P}}$, as explained in the next section, as well as different configurations of Gaussian Process outputs, alongside respective macroeconomic variables, which are discussed in greater detail in Section \ref{sec:Data}. 

%In order to incorporate the unspanned macroeconomic information in the model, in a nonlinear way, we further revise the $\mathbb{P}$ dynamics of $\mathcal{P}_t$. The other assumptions with respect to the model made so far, such as on the short-rate $r_t$ and the $\mathbb{Q}$ dynamics of $\mathcal{P}_t$, remain unchanged.

Our starting point is the $\mathbb{P}$-dynamics for $\mathcal{P}_t$ as in \eqref{PmodelExog} which, for convenience, we simplify to  
\begin{equation}
\label{PmodelExogGP}
s_t = v(M_{t-1}) +\Sigma_{\mathcal{P}} \varepsilon_{t}
\end{equation}
by setting
\begin{equation}\label{P2s}
\mathcal{P}_t - \mu_{\mathcal{P}}^{\mathbb{P}} - \Phi_{\mathcal{P}}^{\mathbb{P}}\mathcal{P}_{t-1}=s_t
\end{equation}
Then we assign a Gaussian Process prior on $v(\cdot)$ in \eqref{PmodelExogGP} and denote it as
\begin{equation}
\pi(v) = GP(v_0,\; K) \label{GPf}
\end{equation}
In the above we assume $v_0=0$, what is commonly done in applications \citep{RasmussenWilliams2006}. Such a prior assumption about $v_0$ is also reasonable in our context, where we are effectively filtering out $GP$s from residuals of a VAR model. Specific choice of kernel $K$ is discussed later. 

At this point it is worth noting that under these assumptions the term $\mu_{\mathcal{P}}^{\mathbb{P}} + v(M_{t-1})$, as backed out from \eqref{PmodelExogGP} and \eqref{P2s}, essentially determines (up to rescaling) the long term mean of the corresponding Ornstein-Uhlenbeck (OU) process that governs $\mathcal{P}_t$. Given that $v(M_{t-1})$ is now a Gaussian Process, the model can be viewed as $N$-dimensional $GP$ mean OU process. Although according to our notation dimension of $v(M_{t-1})$ in \eqref{PmodelExogGP} is $(N\times 1$), we are able to include $G=1,\dots,N$ Gaussian Process outputs in a single model. For example, if $G=2$ and we introduce two $GP$ outputs, one in the second and one in the third equation of \eqref{PmodelExogGP}, then it means that the first element of $v(M_{t-1})$ is effectively zero. It is implemented by setting $K_1$, the first diagonal block of matrix $K$ defined in \eqref{Kcov} below, to zeros. In the linear model, this would be equivalent to restricting the first row of matrix $\Phi_{\mathcal{P}M}^{\mathbb{P}}$ in \eqref{PmodelExogLin} to zeros, what is explained in detail in Appendix \ref{appendix:LinearModel}. %In empirical application that follows, we choose from $G\in\{2,3\}$, whereby in case of $G=2$ we decide to include no GP in the first equation of \eqref{PmodelExogGP}, for reasons named in Section \ref{sec:Data}.

Without loss of generality for $G=3$, we adopt the following vector notation
$$
s_t= \left[\begin{array}{c}
	 s^{(1)}_t \\
	 s^{(2)}_t\\
	 s^{(3)}_t
	\end{array}\right],\;\;
s_{jt}=s^{(j)}_t,\;\;
S_{j}=[s_{j1},\dots,s_{jT}]',\;\;j=1,2,3
$$
$$
v(M_t) = \left[\begin{array}{c}
	v^{(1)}(M_{t}) \\
	v^{(2)}(M_{t})\\
	v^{(3)}(M_{t})
	\end{array}\right],\;\;
v_{jt}=v^{(j)}(M_{t}),\;\;V_{j}=[v_{j0},\dots,v_{jT-1}]',\;\;j=1,2,3
$$
Eventually we consider the concatenated vectors 
\begin{equation}\label{S_F}
S= \left[\begin{array}{c}
	 S_1 \\
	 S_2\\
	 S_3
	\end{array}\right],\;\;V= \left[\begin{array}{c}
	 V_1 \\
	 V_2\\
	 V_3
	\end{array}\right]
\end{equation}
The distribution of $S$ conditional on $V$ is then
\begin{equation}
\label{ymodel} % Y became S, and y became s, since I changed notation  
S|V \sim N\left(V,\;\Sigma_\mathcal{P}\Sigma_\mathcal{P}' \otimes I_T\right)
\end{equation}

The next step is to choose a specific kernel for each $GP$ $v^{(j)}(M_t)$, $j=1,2,3$. We use the same type of kernel for all the $GP$s in the model. Namely, the squared exponential kernel, which is stationary and considered the choice most-widely made to start with in applications \citep{Bishop2006,RasmussenWilliams2006,Murphy2012}. It can be specified as
$$
k_{\ell,\sigma^2}(x_a, x_b) = \sigma^2 \exp \left(-\frac{ \left\Vert x_a - x_b \right\Vert^2}{2\ell^2}\right)
$$
for some scalars (or vectors) $x_a$ and $x_b$, where $\ell$ is called the characteristic length-scale and $\sigma$ is denoted as signal standard deviation. Squared exponential kernel is infinitely differentiable and thus very smooth, what we welcome here since it is the smooth signal within the rough noise that we are after.

To each $V_j$ we thus assign separate yet the same type of kernel $k_{\ell_j,\sigma_j^2}(\cdot,\cdot)$, $j=1,2,3$, what can be expressed as 
$$
K_j=\left[\begin{array}{ccc}
	 k_{\ell_j,\sigma_j^2}(M_0, M_0) & \dots &k_{\ell_j,\sigma_j^2}(M_0, M_{T-1})  \\
	 \vdots & \ddots & \vdots\\
	  k_{\ell_j,\sigma_j^2}(M_{T-1},M_0)& \dots & k_{\ell_j,\sigma_j^2}(M_{T-1}, M_{T-1})
	\end{array}\right],\;\;j=1,2,3
$$
in covariance terms. Following the assumption made about $v_0$ in \eqref{GPf}, it lets us then concretely determine the distribution for each of the $V_j$s as
\begin{equation}\label{Fdist}
V_j \sim N(0_{T\times1},\; K_j),\;\;j=1,2,3
\end{equation}
where $0_{T\times1}$ is a $(T \times 1)$ vector of zeros and $K_j$ is the $(T \times T)$ matrix shown above.

Finally, in order to specify the distribution of $V$, which in our case is a single input $GP$ with multiple outputs, we need to make an assumption about the dependence between $V_j$s. The simplest way, and the one chosen here, is to assume independence between $V_j$s, what implies the following distribution
\begin{equation}
\label{Fmodel}
V\sim N\left(0_{3T\times 1},\; K \right)
\end{equation}
with
\begin{equation}\label{Kcov} % OmegaF became K since I changed notation
K = \left[\begin{array}{ccc}
	 K_1 & 0_T & 0_T  \\
	 0_T & K_2 & 0_T\\
	 0_T & 0_T & K_3
	\end{array}\right]
\end{equation}
which is a $(3T\times 3T)$ matrix, and $0_{3T\times 1}$ and $0_T$ are a $(3T\times 1)$ and $(T\times T)$ vector and matrix of zeros, respectively. We collect the hyper-parameters governing covariance matrix $K$, which are $\sigma_j$ and $\ell_j$, $j=1,2,3$, in $\sigma_K=[\sigma_1,\sigma_2,\sigma_3]'$ and $\ell_K=[\ell_1,\ell_2,\ell_3]'$.

There are also more flexible alternatives for tying up together the $V_j$s and specifying the covariance of $V$, that is $K$, see for example \cite{AlvarezRosascoLawrence2012} for a survey of several methods. These include the intrinsic coregionalisation model, the semiparametric latent factor model and the linear model of coregionalisation, among others. Choosing between such models involves considering trade-offs between flexibility, computational cost and interpretability. Here we adopt \eqref{Kcov} as a convenient starting point for DTSMs. Consequently, such a single input $GP$ with multiple inputs is then equivalent to multiple independent single input $GP$s with a single output.

Next, using standard properties of Gaussian Processes we can proceed to obtain the conditional distribution of $V$ given $S$, the marginal distribution of $S$, by integrating out $V$ from \eqref{ymodel}, and the predictive distribution of $s_{T+1}$, and consequently that of $\mathcal{P}_{T+1}$, given all the data up to time $T$ including $M_T$. For reasons mentioned in Section \ref{sec:predEconomicValue}, in this paper we focus on one-step-ahead predictions. To start with, the marginal distribution of $S$, which \textit{nota bene} stands behind the $\mathbb{P}$-dynamics of $\mathcal{P}_t$, is following
\begin{equation} \label{Smarginal}
S \sim N\left(0_{3T\times 1},\; K+\Sigma_\mathcal{P}\Sigma_\mathcal{P}' \otimes I_T\right)
\end{equation}
To obtain conditional distribution of $V$ given $S$ we observe that joint distribution of $V$ and $S$ is
\begin{equation}\label{FjointS}
\left(\begin{array}{c}
V \\
S
\end{array}
 \right) \sim N\left(\left[\begin{array}{c}0_{3T\times 1} \\
0_{3T\times 1} \end{array}\right],\; \left[\begin{array}{cc}
	 K & K \\
	 K & K_{\mathcal{P}} \\
	\end{array}\right] \right)
\end{equation}
where for clarity of further exposition we defined
\begin{equation}\label{KPcov} % OmegaFP became K_\mathcal{P} since I changed notation
K_{\mathcal{P}}=K+\Sigma_\mathcal{P}\Sigma_\mathcal{P}'\otimes I_T
\end{equation}
Then, by the multivariate normal regression lemma, we get
\begin{equation}\label{FgivenS}
V|S \sim N\left(K K_{\mathcal{P}}^{-1}S,\; K-K K_{\mathcal{P}}^{-1}K \right)
\end{equation}

To arrive at the predictive distribution of $s_{T+1}$, we first derive the joint distribution of $s_{T+1}$ and $S$, using all available macro data, that is $M_{0:T}$. Given the assumption about dependence between $V_j$s, as it is made in \eqref{Kcov}, we first note that for $t=1,\dots,T$
\begin{equation}
Cov(s_t,s_{T+1}) = \left[
					\begin{array}{ccc}
					k_{\ell_1,\sigma_1^2}(M_{t-1}, M_T) & 0 & 0 \\
					0 & k_{\ell_2,\sigma_2^2}(M_{t-1}, M_T) & 0 \\
					0 & 0 & k_{\ell_3,\sigma_3^2}(M_{t-1}, M_T)
					\end{array}
					\right]
\end{equation}
Further, for $j=1,2,3$, we let
$$
k^j_{T+1} = [k_{\ell_j,\sigma_j^2}(M_0, M_T),...,k_{\ell_j,\sigma_j^2}(M_{T-1},M_T)]'
$$ 
to eventually observe that joint distribution of $s_{T+1}$ and $S$ is as follows
\begin{equation}
\left(\begin{array}{c}
s_{T+1} \\
S
\end{array}
 \right) \sim N\left(\left[\begin{array}{c}0_{3\times 1} \\
0_{3T\times 1} \end{array}\right],\; \left[\begin{array}{cc}
	 k^0_{T+1}+\Sigma_\mathcal{P}\Sigma_\mathcal{P}' & k_{T+1}' \\
	 k_{T+1} & K_{\mathcal{P}} \\
	\end{array}\right] \right)
\end{equation}
where $k^0_{T+1}$ and $k_{T+1}$, which are $(3\times 3)$ and $(3T\times 3)$ matrices, respectively, are specified as
\begin{equation}\label{k_0_t1}
k^0_{T+1} = \left[\begin{array}{ccc} k_{\ell_1,\sigma_1^2}(M_T,M_T) & 0 & 0 \\
	0 & k_{\ell_2,\sigma_2^2}(M_T,M_T) & 0 \\
	0 & 0 & k_{\ell_3,\sigma_3^2}(M_T,M_T) \\
	\end{array}\right]
\end{equation}
and
\begin{equation}\label{k_t1}
k_{T+1} = \left[\begin{array}{ccc} k^1_{T+1} & 0_{T\times 1} & 0_{T\times 1} \\
	0_{T\times 1} & k^2_{T+1} & 0_{T\times 1} \\
	0_{T\times 1} & 0_{T\times 1} & k^3_{T+1} \\
	\end{array}\right]
\end{equation}
Then, again by the multivariate normal regression lemma, we get the predictive distribution of $s_{T+1}$ given all the data up to time $T$ including $M_T$
\begin{equation}
s_{T+1}|S,M_{0:T} \sim N\left(k_{T+1}'K_{\mathcal{P}}^{-1}S,\; k^0_{T+1}+ \Sigma_\mathcal{P}\Sigma_\mathcal{P}' -k_{T+1}'K_{\mathcal{P}}^{-1}k_{T+1}\right)
\end{equation}
what translates into the corresponding predictive distribution of $\mathcal{P}_{T+1}$ as follows
\begin{equation}
\mathcal{P}_{T+1}|\mathcal{P}_{T},S,M_{0:T} \sim N\left(\mu_{\mathcal{P}}^{\mathbb{P}} + \Phi_{\mathcal{P}}^{\mathbb{P}}\mathcal{P}_{T}+k_{T+1}'K_{\mathcal{P}}^{-1}S,\; k^0_{T+1}+\Sigma_\mathcal{P}\Sigma_\mathcal{P}'-k_{T+1}'K_{\mathcal{P}}^{-1}k_{T+1}\right)
\end{equation}

\subsection{Likelihood and Risk Price Restrictions}%, and the Gaussian Process}
\label{subsec:likelihoodGaussian}

Statistical inference can be performed using the observations $Y=\{y_t,\mathcal{P}_{t}: t=0,1,\dots,T\}$ and $M=\{M_t: t=0,1,\dots,T-1\}$. The likelihood factorizes into two parts stemming from the $\mathbb{P}$ and $\mathbb{Q}$ respectively. For $N$ observable factors, the joint likelihood (conditional on the initial point $\mathcal{P}_{0}$) can now be written as
\begin{equation}\label{likelihood}
\begin{array}{l}
f(Y|M,\theta,\widehat{\sigma}_K)=\left\{\prod_{t=0}^T f^{\mathbb{Q}}(y_{t}|\mathcal{P}_{t}, k_{\infty}^{\mathbb{Q}}, g^{\mathbb{Q}}, \Sigma_{\mathcal{P}}, \sigma_{e}^{2})\right\} \times \\ 
\;\;\;\;\;\;\;\;\;\;\;\;\;\;\;\;\;\;\;\;\;\;\;\;\;\; \left\{\prod_{t=1}^T f^{\mathbb{P}}(\mathcal{P}_{t}|\mathcal{P}_{t-1},M_{t-1}, k_{\infty}^{\mathbb{Q}}, g^{\mathbb{Q}},\lambda_{0\mathcal{P}},\lambda_{1\mathcal{P}}, \Sigma_{\mathcal{P}},\ell_K,\widehat{\sigma}_K)\right\}
\end{array}
\end{equation}
where the $\mathbb{Q}$-likelihood components $f^{\mathbb{Q}}(\cdot)$ are given by \eqref{Qmodel} and capture the cross-sectional dynamics of the risk factors and the yields, whereas $\mathbb{P}$-likelihood components $f^{\mathbb{P}}(\cdot)$ are obtained from \eqref{loglikP} below and capture the time-series dynamics of the observed risk factors. The parameter vector is set to $\theta=(\sigma_{e}^{2},k_{\infty}^{\mathbb{Q}}, g^{\mathbb{Q}},\lambda_{0\mathcal{P}},\lambda_{1\mathcal{P}},\Sigma_{\mathcal{P}}, \ell_K)$ %To assure identification, 
and we tune $\sigma_K$ %is calibrated 
in-sample to fix it out-of-sample at $\widehat{\sigma}_K$. Details of %its calibration
the tuning procedure are in Appendix \ref{appendix:GPcalibration}.

For brevity of further exposition, we let $\lambda^{\mathcal{P}}=[\lambda_{0\mathcal{P}},\lambda_{1\mathcal{P}}]$ and $\lambda=\lambda_{1\mathcal{P}}$. If all the entries in $\lambda^{\mathcal{P}}$ are free parameters we get the maximally flexible model (model $M_0$ in this paper). Alternative specifications, with some of these entries set to zero, have been proposed in the literature. For details, see the associated discussion with related references in \cite{DTKK2021b}. Overall, in most models the set of unrestricted parameters is usually a subset of $\lambda^{\mathcal{P}}$. In this paper, we use restriction set with optimal model performance, in particular with respect to economic value, as evidenced in \cite{DTKK2021b}. It is also the same restriction set as in \cite{DTKK2021a}, for reasons mentioned therein. Namely, we only leave $\lambda_{1,2}$ unrestricted, as it is in model $M_1$, here and in the other papers.
%set with optimal predictive performance, understood predominantly as economic value and less as statistical predictability, among all possible restriction sets, as evidenced in \cite{DTKK2021b}, based on out-of-sample period of data closely overlapping ours, see Section \ref{sec:Data}. \cite{DTKK2021b} develop a sequential version of stochastic search variable selection scheme and arrive at a conclusion that restriction set offering such an optimal predictive performance based on out-of-sample period from 2007 to 2016 is to only leave $\lambda_{1,2}$ unrestricted (model $M_1$ there and in this paper). In what follows, we would like to build up further on this result. Our choice is conceptually close to \cite{Bauer18} who suggests using Bayesian model choice, aiming to maximize model evidence of each restriction specification. Models that are optimal in the Bayesian sense, i.e. achieving the highest model evidence, are typically parsimonious and therefore exhibit good predictive performance. It is also to some extent in accordance with the early work on risk dynamics, such as \cite{Duffee02}, which finds that variation in the price of level risk is necessary to capture the failure of the Expectations Hypothesis (EH). This variation was usually captured by linking the price of level risk to the slope of the term structure \citep{Duffee11}.

Consequently, the likelihood specification of \eqref{likelihood} can now be restated as
\begin{equation}\label{likelihoodlambda12}
\begin{array}{l}
f(Y|M,\theta,\widehat{\sigma}_K)=\left\{\prod_{t=0}^T f^{\mathbb{Q}}(y_{t}|\mathcal{P}_{t}, k_{\infty}^{\mathbb{Q}}, g^{\mathbb{Q}}, \Sigma_{\mathcal{P}}, \sigma_{e}^{2})\right\} \times \\ 
\;\;\;\;\;\;\;\;\;\;\;\;\;\;\;\;\;\;\;\;\;\;\;\;\;\; \left\{\prod_{t=1}^T f^{\mathbb{P}}(\mathcal{P}_{t}|\mathcal{P}_{t-1},M_{t-1}, k_{\infty}^{\mathbb{Q}}, g^{\mathbb{Q}},\Sigma_{\mathcal{P}},\ell_K,\lambda_{1,2},\widehat{\sigma}_K)\right\}
\end{array}
\end{equation}
where $\theta=(\sigma_{e}^{2},k_{\infty}^{\mathbb{Q}}, g^{\mathbb{Q}},\Sigma_{\mathcal{P}}, \ell_K,\lambda_{1,2})$ is revised accordingly. To obtain $\mathbb{P}$-likelihood components $f^{\mathbb{P}}(\cdot)$ above, we resort to Gaussian Process formulation established so far, and exploit the associated marginal distribution of $S$ in \eqref{Smarginal}, together with \eqref{KPcov}, to eventually arrive at the standard log-likelihood representation of $\left\{\prod_{t=1}^T f^{\mathbb{P}}(\cdot)\right\}$ in \eqref{likelihoodlambda12}, namely
\begin{equation}
\log{\left\{\prod_{t=1}^T f^{\mathbb{P}}(\cdot)\right\}} =  -\frac{T N}{2}\log{2\pi} - \frac{1}{2} \log |K_\mathcal{P}|  - \frac{1}{2} \| K_\mathcal{P}^{-\frac{1}{2}} S \|^{2}
\label{loglikP}
\end{equation}
where $|\cdot|$ is matrix determinant and $\|\cdot\|^2$ denotes Euclidean norm squared.
 
\section{Sequential Estimation, Learning, and Forecasting}
\label{sec:Sequential}

%In this section we develop a sequential Monte Carlo (SMC) framework for Gaussian DTSMs with unspanned nonlinear macros. We draw from the work of \cite{Chopin02,Chopin04} (see also \cite{DelMoral06}), and make the necessary adaptations to tailor the methodology to the data and models considered in this paper. Furthermore, we extend the framework to allow for sequential Bayesian treatment of exogenous macroeconomic information both in a nonlinear and in a linear manner. The former using Gaussian Processes. Overall, the developed framework allows the efficient performance of tasks such as sequential parameter estimation and forecasting. We begin by providing the main skeleton of the scheme and then explain the details of its specific parts, such as the MCMC scheme including Gaussian Processes, and the framework for obtaining and evaluating the economic value of forecasts.

\subsection{Sequential Framework with Gaussian Processes}
\label{subsec:IBIS}

Let $Y_{0:t}=(Y_0,Y_1\dots,Y_t)$ denote all bond related data available up to time $t$, such that $Y_{0:T}=Y$, and $M_{0:t-1}=(M_0,M_1\dots,M_{t-1})$ denote all macro data available up to time $t-1$, such that $M_{0:T-1}=M$. Similarly, the likelihood based on data up to time $t$ is $f(Y_{0:t}|M_{0:t-1},\theta,\widehat{\sigma}_K)$ and is defined in \eqref{likelihood}. Combined with a prior on the parameters $\pi(\theta)$, see the Appendix \ref{appendix:priors} for details, it yields the corresponding posterior 
\begin{equation}\label{posterior}
 \pi(\theta|Y_{0:t},M_{0:t-1},\widehat{\sigma}_K)= \frac{1}{m(Y_{0:t}|M_{0:t-1},\widehat{\sigma}_K)}f(Y_{0:t}|M_{0:t-1},\theta,\widehat{\sigma}_K)\pi(\theta)
\end{equation} 
where $m(Y_{0:t}|M_{0:t-1},\widehat{\sigma}_K)$ is the model evidence based on data up to time t. Moreover, the posterior predictive distribution, which is the main tool for Bayesian forecasting, is defined as
\begin{equation}\label{predictive}
f(Y_{t+h}|Y_{0:t},M_{0:t},\widehat{\sigma}_K)=\int f(Y_{t+h}|Y_{0:t},M_{0:t},\theta,\widehat{\sigma}_K)\pi(\theta|Y_{0:t},M_{0:t-1},\widehat{\sigma}_K)d\theta
\end{equation}
where $h$ is the prediction horizon. As explained further, in Section \ref{sec:predEconomicValue}, in this paper we focus on $h=1$. Due to the Gaussian Process formulation, predictions depend computationally on all available data, what is reflected by conditioning on $Y_{0:t}$ and $M_{0:t}$ in the first term under the integral in \eqref{predictive}. In theory, it is a potential issue as computational cost of deriving the predictive distribution grows fast. In particular, as new data arrives it becomes more costly to invert covariance matrix in \eqref{KPcov}. However, given that time series we consider are relatively short, computational time is far from prohibitive. Therefore, it is not a practical concern.

Note also that the predictive distribution in \eqref{predictive} incorporates parameter uncertainty by integrating $\theta$ out according to the posterior in \eqref{posterior}. Usually, prediction is carried out by expectations with respect to \eqref{predictive}, e.g. $E(Y_{t+h}|Y_{0:t},M_{0:t},\widehat{\sigma}_K)$ but, since \eqref{predictive} is usually not available in closed form, Monte Carlo can be applied in the presence of samples from $\pi(\theta|Y_{0:t},M_{0:t-1},\widehat{\sigma}_K)$. This process may facilitate various forecasting tasks; for example forecasting several points, functions thereof, and potentially further ahead in the future. A typical forecasting evaluation exercise requires taking all the consecutive times $t$ from the nearest integer of, say, $T/2$ to $T-1$. In each of these times, $Y_{0:t}$ serves as the training sample, and points of $Y$ after $t$ are used to evaluate the predictions. Hence, performing such a task requires samples from \eqref{predictive}, and thus from $\pi(\theta|Y_{0:t},M_{0:t-1},\widehat{\sigma}_K)$, for several times $t$. Note that this procedure can be quite intensive and in some cases not feasible.

Another approach that can also handle forecasting assessment tasks is to use sequential Monte Carlo (see, \cite{Chopin02} and \cite{DelMoral06}) to sample from the sequence of distributions $\pi(\theta|Y_{0:t},M_{0:t-1},\widehat{\sigma}_K)$ for $t=0,1,\dots,T$. A general overview of the Iterated Batch Importance Sampling (IBIS) scheme of \cite{Chopin02}, see also \cite{DelMoral06} for a more general framework, is provided in Algorithm \ref{tab:ibis}.
\begin{algorithm}[!ht]
\begin{flushleft}
%\hrule
%\medskip
{\itshape \small
\vspace{0.2cm}
Initialize $N_{\theta}$ particles by drawing independently $\theta_{i}\sim \pi(\theta)$ with importance weights $\omega_{i}=1$, $i=1,\dots,N_{\theta}$. For $t,\dots,T$ and each time for all $i$:\vspace{0.2cm}
\begin{itemize}
\item[(a)] Calculate the incremental weights from
$$
u_t(\theta_{i},Y_{0:t-1},M_{0:t-1}) =f(Y_{t}|Y_{0:t-1},M_{0:t-1},\theta_{i},\widehat{\sigma}_K)
$$
where the Gaussian Process formulation makes them computationally dependant on all available data, what is reflected by conditioning on $Y_{0:t-1}$ and $M_{0:t-1}$.
\item[(b)] Update the importance weights $\omega_{i}$ to $\omega_{i}u_t(\theta_{i},Y_{0:t-1},M_{0:t-1})$.
\item[(c)] If some degeneracy criterion (e.g. ESS($\omega$)) is triggered, perform the following two sub-steps:
\begin{itemize}
\item[ (i)] Resampling: Sample with replacement $N_{\theta}$ times from the set of $\theta_{i}$s according to their weights $\omega_i$. The weights are then reset to one.
\item[(ii)] Jittering: Replace $\theta_i$s with $\tilde{\theta}_i$s by running MCMC chains with each $\theta_i$ as input and $\tilde{\theta}_i$ as output. Set $\theta_i=\tilde{\theta}_i$.
\end{itemize} 
\end{itemize}}
%\medskip
%\hrule
\end{flushleft}
\caption{IBIS algorithm for Gaussian Affine Term Structure Models with unspanned nonlinear macros}
\label{tab:ibis}
\end{algorithm}
The degeneracy criterion is typically defined through the Effective Sample Size ($ESS$) which equals to
\begin{equation} \label{ESS}
ESS(\omega)=\frac{(\sum_{i=1}^{N_{\theta}} \omega_{i})^2}{\sum_{i=1}^{N_{\theta}} \omega_{i}^2}
\end{equation}
and is of the form $ESS(\omega)<\alpha N_{\theta}$ for some $\alpha \in (0,1)$, where $\omega$ is the vector containing the weights. 

The IBIS algorithm provides a set of weighted $\theta$ samples, called also particles, that can be used to compute expectations with respect to the posterior distribution, $E[g(\theta)|Y_{0:t},M_{0:t-1},\widehat{\sigma}_K]$, for all $t$ using the estimator $\sum_{i}[\omega_i g(\theta_{i})]/\sum_{i}\omega_i$. \cite{Chopin04} shows consistency and asymptotic normality of this estimator as $N_{\theta}\rightarrow \infty$ for all appropriately integrable $g(\cdot)$. The same holds for expectations with respect to the posterior predictive distribution, $f(Y_{t+h}|Y_{0:t},M_{0:t},\widehat{\sigma}_K)$. The weighted $\theta$ samples can be conveniently transformed into weighted samples from $f(Y_{t+h}|Y_{0:t},M_{0:t},\widehat{\sigma}_K)$ by just applying $f(Y_{t+h}|Y_{0:t},M_{0:t},\theta,\widehat{\sigma}_K)$. A very useful by-product of the IBIS algorithm is the ability to compute 
$$
m(Y_{0:t}|M_{0:t-1},\widehat{\sigma}_K)=f(Y_{0:t}|M_{0:t-1},\widehat{\sigma}_K)
$$
which is the criterion for conducting formal Bayesian model choice. Computing the following quantity in step (a) in Algorithm \ref{tab:ibis} yields a consistent and asymptotically normal estimator of $f(Y_t|Y_{0:t-1},M_{0:t-1},\widehat{\sigma}_K)$, namely
\begin{equation}
m_t = \frac{1}{\sum_{i=1}^{N_{\theta}}\omega_{i}}\sum_{i=1}^{N_{\theta}}\omega_{i}u_{t}(\theta_{i},Y_{0:t-1},M_{0:t-1})
\end{equation}
An additional advantage of SMC is that it provides an alternative when MCMC algorithms are not mixing well and their convergence properties are poor. Generally, it is more robust when the target posterior is problematic, e.g. multimodal. %Finally, as we demonstrate in Section \ref{sec:application}, the sequential nature of the algorithm allows it to produce informative descriptive output to monitor the evolution of key parameters in time.

So as to apply the IBIS output to models and data in this paper, the following adaptations and extensions are implemented. Similar to \cite{DTKK2021b}, and pursuing motivation therein, we pool together the advantages of data tempering and adaptive tempering \citep{jasra2011,Schafer13,Kantas2014} in a
%First, the choice of defining the incremental weights in step (a) in Table \ref{tab:ibis}, also known as data tempering, is suitable for getting access to sequences of predictive distributions, needed to assess forecasting performance, but at the same time it is quite prone to numerical stability issues and very low effective sample sizes, in particular early on, that is at the initial time points. This is because the learning rate is typically higher at the beginning, especially when transitioning from a vague prior. An alternative approach that guarantees a pre-specified minimum effective sample size level, and therefore some control over the Monte Carlo error, is to use adaptive tempering; see, for example, \cite{jasra2011}. In order to combine the benefits of both approaches we use a 
hybrid adaptive tempering scheme which we outline in Appendix \ref{appendix:adaptivetempering}. %The idea of this scheme is to use adaptive tempering within each transition between the posteriors based on $(Y_{0:t},M_{0:t-1},\widehat{\sigma}_K)$ and $(Y_{0:t+1},M_{0:t},\widehat{\sigma}_K)$ for each $t$. Similar ideas have been applied in \cite{Schafer13} and \cite{Kantas2014}. 
Because the MCMC algorithm used here is an extended version of \cite{Bauer18} and so it consists of independence samplers that are known to be unstable, we exploit the IBIS output and estimate posterior moments to arrive at independence sampler proposals; see Appendix \ref{appendix:MCMC} for details, and \cite{DTKK2021b} for additional rationale. Under such amended framework, and quite crucially in this paper, we extend the framework presented in Sections \ref{subsec:GPOUModel} and \ref{subsec:likelihoodGaussian} to handle Gaussian Processes sequentially. %Third, we note that the MCMC sampler, used in sub-step (ii) of step (c) in Table \ref{tab:ibis}, needs to be automated as it will have to be rerun for each time point and particle without the luxury of having initial trial runs, as it is often the case when running a simple MCMC on all the data. The problem is intensified by the fact that 
In the end, we apply IBIS output in the optimization of a model-driven dynamically rebalanced portfolio of bond excess returns and inspect its economic value, following in the footsteps of \cite{DTKK2021b}.

In empirical work, we use $N_{\theta}=2000$ particles, $5$ MCMC steps when jittering, and with regards to minimum $ESS$ we set $\alpha=0.7$. The choice of $5$ steps at the jittering stage is led by quite well mixing behaviour of the underlying MCMC. We monitored the correlation between particles before and after that stage to realize that performance was already reasonable with this number of iterations.

\subsection{Assessing Predictive Performance and Economic Value}
\label{sec:predEconomicValue}

Leveraging the evaluation framework summarized in this section, we seek to understand whether macroeconomic information introduced into Gaussian ATSMs in a nonlinear manner using Gaussian Processes lets us predict excess returns better than when macros enter these models linearly. Furthermore, we attempt to explore whether such statistical predictability, if any, can be turned into consistent economic benefits for bond investors.

As explained further in Section \ref{subsec:yieldsmacros}, we use lagged macro data at monthly frequency. Given the limitations to availability of macroeconomic information going forward and its exogenous and not endogenous nature we assume in this paper, what prevents us from associated forecasting further than a month ahead, we concentrate on 1-month prediction horizon. Thus, we refrain from making any assumptions about forward paths of the underlying macros. Consequently, returns we consider in our analysis are non-overlapping.

% \ \\
% \textbf{TDT:} Here elaborate more on the problem statement in relation to nonlinear vs linear treatment of macros, citing respective literature>
% %Reconsider whether not to make a random walk assumption about macros and still show results for different $h$. If so, then include log predictive score again (too late for this now!)
% \ \\

% \subsubsection{Bond Excess Returns}
% \label{sec:R2OS}

% Define the observed $h$-holding period return from buying an $n$-year bond at time $t$ and selling it at time ($t+h$) as
% %
% \begin{equation}\label{holding}
% r_{t,t+h}^{n} = p_{t+h}^{n-h} - p_{t}^{n}
% \end{equation}
% %
% where $p_{t+h}^{n-h}$ is the log price of the ($n-h$)-period bond at time $(t+h)$ and $p_{t}^{n}$ is the log price of the $n$-period bond at time $t$. The latter translates to the corresponding yield in the following manner
% %
% \begin{equation}\label{price2yield}
% y_{t}^{n} = -\frac{1}{n}p_{t}^{n} 
% \end{equation}
% %
% Furthermore,

Along the lines of \cite{DTKK2021b}, yet with $h=1$ in mind, we define the observed continuously compounded excess return of an $n$-year bond as the difference between the holding period return of the $n$-year bond and the $h$-period yield as
\begin{equation}\label{excess}
rx_{t,t+h}^{n} = -(n-h) y_{t+h}^{n-h} + ny_{t}^{n} - hy_{t}^{h}
\end{equation}
If, instead of taking the observed one, we take the model-implied continuously compounded yield $y_{t}^{n}$, calculated according to (\ref{newyield}), we arrive at the predicted excess return $\widetilde{rx}_{t,t+h}^{n}$ which becomes
\begin{equation}\label{fitted} 
\widetilde{rx}_{t,t+h}^{n} 
%&= -(n-h) \tilde{y}_{t+h}^{n-h} + n\tilde{y}_{t}^{n} - h\tilde{y}_{t}^{h} \\
= A_{n-h,\mathcal{P}} - A_{n,\mathcal{P}} + A_{h,\mathcal{P}} + B'_{n-h,\mathcal{P}} \widetilde{\mathcal{P}}_{t+h} - (B_{n,\mathcal{P}}-B_{h,\mathcal{P}})'\mathcal{P}_{t}
\end{equation}
where $\mathcal{P}_{t}$ is observed and $\widetilde{\mathcal{P}}_{t+h}$ is a prediction from the model. Our developed framework, see Algorithm \ref{tab:ibis}, allows drawing from the predictive distribution of $(\widetilde{\mathcal{P}}_{t+h},\widetilde{rx}_{t,t+h}^{n})$ based on all information available up to time $t$. More specifically, for each $\theta_i$ particle the $\mathbb{P}$-dynamics of $\mathcal{P}_t$ can be used to obtain a particle of $\widetilde{\mathcal{P}}_{t+h}$, which then can be transformed into a particle of $\widetilde{rx}_{t,t+h}^{n}$ via equation \eqref{fitted}. Detailed steps are outlined in Algorithm \ref{tab:predictive}.
\begin{algorithm}[!ht]
\begin{flushleft}
%\hrule
%\medskip
{\itshape \small
\vspace{0.2cm}
First, at time $t$, for some $n$ and $h=1$, using $(\omega_i,\theta_i)$, $i=1,...,N_{\theta}$, from IBIS algorithm, iterate over $i$:\vspace{0.2cm}
\begin{itemize}
\item[(a)] Given $\theta_i$, compute $A_{i_1,\mathcal{P}}$ and $B_{i_1,\mathcal{P}}$, for $i_1\in\{1,n-1,n\}$, from \eqref{rotation2} and \eqref{rotation2_2}. 

\item[(b)] Given $\theta_i$, obtain prediction of $\mathcal{P}_{t+1}$ by drawing from 
$$
\widetilde{\mathcal{P}}_{t+1}^{(i)}|\mathcal{P}_{t},S,M_{0:t} \sim N\left(\mu_{\mathcal{P}}^{\mathbb{P}} + \Phi_{\mathcal{P}}^{\mathbb{P}}\mathcal{P}_{t}+k_{t+1}'K_{\mathcal{P}}^{-1}S,\; k^0_{t+1}+\Sigma_\mathcal{P}\Sigma_\mathcal{P}'-k_{t+1}'K_{\mathcal{P}}^{-1}k_{t+1}\right)
$$
where $S$, $K_{\mathcal{P}}$, $k_{t+1}^0$ and $k_{t+1}$ are defined in \eqref{S_F}, \eqref{KPcov}, \eqref{k_0_t1} and \eqref{k_t1}, respectively, and except for $S$ they all depend on $\widehat{\sigma}_K$.

\item[(c)] Compute particle prediction of $rx_{t,t+1}^{n}$ as
$$
\widetilde{rx}_{t,t+1}^{n(i)} = A_{n-1,\mathcal{P}} - A_{n,\mathcal{P}} + A_{1,\mathcal{P}} + B'_{n-1,\mathcal{P}} \widetilde{\mathcal{P}}_{t+1}^{(i)} - (B_{n,\mathcal{P}}-B_{1,\mathcal{P}})'\mathcal{P}_{t}
$$
\end{itemize}

Second, since $(\omega_i,\widetilde{\mathcal{P}}_{t+1}^{(i)},\widetilde{rx}_{t,t+1}^{n(i)})$, $i=1,...,N_\theta$, is a particle approximation to predictive distribution of $(\mathcal{P}_{t+1},rx_{t,t+1}^{n})$, compute point prediction of $rx_{t,t+1}^{n}$ using particle weights $\omega_i$ as
$$
\widetilde{rx}_{t,t+1}^{n}=\frac{1}{\sum_{i=1}^{N_\theta}\omega_{i}}\sum_{i=1}^{N_{\theta}}\omega_{i}\widetilde{rx}_{t,t+1}^{n(i)}
$$

Third, repeat above two steps for different $n$.}
%\medskip
%\hrule
\end{flushleft}
\caption{Predictive distribution of excess returns for Gaussian Affine Term Structure Models with unspanned nonlinear macros}
\label{tab:predictive}
\end{algorithm}

To assess the predictive ability of the models considered, we compute the associated out-of-sample $R^{2}$ ($R_{os}^{2}$), due to \cite{Campbell08}. To measure the resulting economic value generated by each model, we consider a Bayesian investor with power utility preferences. Our Bayesian learner solves an
asset allocation problem getting optimal portfolio weights which we use to compute the $CER$ as in \cite{Johannes14} and \cite{Gargano19}. To that end, we follow the exposition in \cite{DTKK2021a}, where we refer the reader for further details. Similar to the latter paper, we adopt the Expectations Hypothesis (EH) as initial empirical benchmark, however instead of then looking at the performance relative to model $M_1$ \citep{DTKK2021b}, we compare the nonlinear with the corresponding linear models. Again, we only focus on a 1-month prediction horizon.% ($h=1$).

\section{Data and Models}
\label{sec:Data}

%In this section we discuss the data used in the paper. %Specifically, we elaborate on the US Treasury %yields data we employ in this paper, alongside the macroeconomic variables we introduce in our %analysis as well. We provide arguments behind the way we split them into a training and a testing %subsample. 
%We also explain what models we consider, as distinguished by different positions unspanned, nonlinear %or linear, macros take in the given model. Eventually, we describe how we %arrive at such choice, in %particular based on some rationale behind 
%use risk premium factor, rationale behind which is provided in \cite{Duffee11}, to inspect whether information coming from macroeconomic variables we consider, in particular their components which are hidden from the yield curve, is \textit{de facto} nonlinear in nature.

\subsection{Yields and Macros}
\label{subsec:yieldsmacros}

In terms of bond yields, we analyze the same data as these described in \cite{DTKK2021a}, splitting them in the same training (in-sample) and testing (out-of-sample) periods, thus for further details we refer the reader therein.

When it comes to macroeconomic information about the US, we consider %three
two variables which are well covered in the literature. These include, core\footnote{In contrast to \cite{Cieslak15}, who devise trend inflation by appropriately smoothing core inflation, we use the latter as is to arrive at our own, nonlinear function thereof.
%Estimating potentially nonlinear functions thereof using Gaussian Processes with squared exponential kernel, as we do here, resembles a smoothing process itself. We thus avoid such overlay.
} inflation ($CPI$) as in \cite{Cieslak15}, as well as the three-month
moving average of the Chicago Fed National Activity Index ($GRO$) from \cite{Joslin14}, which is a measure of current economic conditions.
%, as well as one variable from \cite{Ludvigson09}, namely
%real activity ($F1$) and 
%the stock market factor ($F8$).
Together they offer parsimonious yet comprehensive enough, for our purposes, description of the US economy.
%factors. Moreover, we complement these with unemployment rate ($UNR$) and manufacturing capacity utilization ($MNF$) we take from St. Louis FRED (https://fred.stlouisfed.org). 
%All
Both macro variables are at a monthly frequency. They are also seasonally adjusted and revised. Unlike $GRO$, which is a smoothed %and $F8$, an 
index %and a factor respectively, which are 
in levels, $CPI$ is in percent changes from year ago. The underlying data period is such that it corresponds to this for yields, as discussed above, and also reflects the lagged character assumed for the macros in all models we consider in this paper. Importantly, we standardize the %all
macros only in the nonlinear case and, to that end, we use individual means and standard deviations calculated in-sample for the out-of-sample operations, to avoid look-ahead bias.   

\subsection{Models and Rationale Behind}
\label{subsec:modelsrationale}

In terms of models, we consider several alternatives when it comes to positions the unspanned, nonlinear or linear, macros take in the given model. More specifically, in the nonlinear case these are
\begin{equation}\label{GPspec}
GP_{ijk}(M)\;\;\; i,j,k\in\{0,1\}
\end{equation}
and for the linear model we replace $GP$ with $LM$ in this notation. In the above, $M$ refers to a specific macroeconomic variable, $CPI$ or $GRO$, introduced in the particular model. For example, for $i=0$ and $j=k=1$, that is $GP_{011}(M)$, index $011$ means that macroeconomic impact is allowed in the second and third equation in \eqref{PmodelExog}, under the assumption that $N=3$, $G=2$ and $R=1$. We only investigate a subset of the available alternative models in greater detail.
%The rationale behind why we set the third element in this index equal to 1 at all times is that unspanned information in this particular equation is decisive for predictive and economic performance of the model, at least by judging when latent factors are pursued to that end \citep{DTKK2021a}. We expect this to be the case for unspanned nonlinear macros as well. Otherwise, we rather let the model parameters and the data speak when it comes to deciding whether certain macros we use affect the other principal components of the yield curve or not.
%We also %definitely
Investigating and determining which macros, beyond the two included in our analysis, represent best match for specific PCs of the US yield curve, when interacting with the latter in a nonlinear or linear manner, is left for future research. %Therefore, and for reasons mentioned next, to limited extent only we alternate between 0 and 1 when it comes to its first and second elements.

As result of risk price restrictions (on $\lambda_{\mathcal{P}}$) we adopt in this paper from \cite{DTKK2021b}, without loss of generality for $i=j=k=1$ in \eqref{GPspec}, we define the risk premium factor as
\begin{equation}\label{RPF111}
RP^V_t(M_{t-1}) = \lambda\mathcal{P}_t + v(M_{t-1}) = \left[\begin{array}{c}
\lambda_{1,2}\mathcal{P}_{2,t} + v^{(1)}(M_{t-1}) \\
v^{(2)}(M_{t-1}) \\
v^{(3)}(M_{t-1})
\end{array}\right]
\end{equation}
where the first element is similar to \cite{Duffee11}, however in our case its time-varying component, $RP_t=\lambda_{1,2}\mathcal{P}_{2,t}$, is a restricted version (only $\lambda_{1,2}$ is free) of the risk premium factor defined in Duffee's paper. The component $v^{(1)}(M_{t-1})$ is in that particular case the first element of the unspanned nonlinear macro $v(M_{t-1})$. If instead $i=0$ and $j=k=1$ in \eqref{GPspec}, then
\begin{equation}\label{RPF011}
RP^V_t(M_{t-1}) = \left[\begin{array}{c}
\lambda_{1,2}\mathcal{P}_{2,t} \\
v^{(2)}(M_{t-1}) \\
v^{(3)}(M_{t-1})
\end{array}\right]
\end{equation}
where the first element, which is related to level risk, equals $RP_t$ and is also equivalent to the corresponding factor in model $M_1$ by \cite{DTKK2021b} we adopt risk price restrictions from.

Following from \cite{Duffee11}, the time-varying risk premium factor $RP_t$, which in the case therein is a linear combination of the state vector obtained from model shocks to yields, determines the compensation investors require to face fixed-income risk from $t$ to $t+1$, and it contains all information relevant to predicting one-step-ahead, yet not $h$-step-ahead for $h>1$, excess returns. We instead assume that investors require compensation for level risk stemming from changes to slope only. However, the term $v(M_{t-1})$ in \eqref{RPF111}, which is \textit{a priori} unspanned and potentially nonlinear function ($v$) of the underlying macroeconomic variable ($M$), affects this compensation. It is important to understand, if 
%(and when yes then by how much?)
this impact is not distorting information relevant to predicting excess returns, already available from $RP_t$ alone. One way to achieve this goal is to perform predictability and economic value exercises for the models considered and make appropriate comparisons. Thus, we do so in Section \ref{sec:application}.

Different from \cite{Duffee11}, we obtain the state vector from observed yields directly. Nevertheless, for $i,j,k \in \{0,1\}$ in \eqref{GPspec}, we are able to similarly define the hidden part of the risk premium factor, as the part unspanned by $\mathcal{P}_t$, in the following way
\begin{align}
\widetilde{RP}^V_t(M_{t-1}) &= RP^V_t(M_{t-1}) - E\left[RP^V_t(M_{t-1})|\mathcal{P}_t\right] \\
&= v(M_{t-1}) - E[v(M_{t-1})|\mathcal{P}_t] \label{RPFhiddenshort} 
\end{align}
where $E[v(M_{t-1})|\mathcal{P}_t]$ is the projection of unspanned nonlinear macros on principal components obtained from observed yields, which is thus spanned by these PCs, and as such not hidden from the yield curve. After \cite{Duffee11}, the hidden $\widetilde{RP}^V_t(M_{t-1})$ can be estimated as residual from a regression of the former on the latter and expressed as
\begin{equation}\label{RPFreg}
\widetilde{RP}^V_t(M_{t-1}) \equiv v(M_{t-1}) - a - b'\mathcal{P}_t 
\end{equation}
where we take $v(M_{t-1})$ as the mean from its posterior distribution and $a$ and $b$ are the underlying Ordinary Least Squares parameter estimates. We also note that
$$
E[v(M_{t-1})|\mathcal{P}_t] = a + b'\mathcal{P}_t
$$
and, by rearranging terms in \eqref{RPFreg}, arrive at the following decomposition
\begin{equation}
v(M_{t-1}) \equiv E[v(M_{t-1})|\mathcal{P}_t] + \widetilde{RP}^V_t(M_{t-1}) \label{vDecomposition} 
\end{equation}
of unspanned nonlinear macros $v(M_{t-1})$, into two orthogonal components. While the first, $E[v(M_{t-1})|\mathcal{P}_t]$, which is an affine transformation of the PCs, is by definition not hidden from the yield curve, the second, $\widetilde{RP}^V_t(M_{t-1})$, duly is.

Then we can examine to what extent $v(M_{t-1})$ assumed nonlinear \textit{a priori} is actually such \textit{a posteriori}. More importantly, however, we investigate whether the same can be stated about its hidden component $\widetilde{RP}^V_t(M_{t-1})$. To that end, we separately regress these, as well as $E[v(M_{t-1})|\mathcal{P}_t]$, on the underlying macroeconomic variable $M_{t-1}$ and inspect the resulting explanatory powers behind such linear associations, see %Tables \ref{table:R2adjRPFEFPFM001010110} and \ref{table:R2adjRPFEFPFM011111}. , see 
Section \ref{sec:RPFandHiddenNonlin} for related discussion. By doing so, we can determine whether there is any connection especially between the hidden component of the risk premium factor being nonlinear and the predictability of and economic benefits from nonlinear models we consider in this paper. In what follows, we eventually demonstrate that, when the latter outperform the linear ones on these two important fronts, this is indeed the case.

\section{Empirical Results}
\label{sec:application}

\subsection{Uncovering Nonlinearities}
\label{subsec:YCandNonlinDyn}

The data contain an important economic event and thus it is interesting to have a look at the plots of individual elements in $v(M_{t-1})$ against $M_{t-1}$, for $CPI$ and $GRO$, to see how it affects our analysis. These plots are generated from the IBIS output, as described in Section \ref{subsec:IBIS} and further in Appendix \ref{appendix:adaptivetempering}. In particular, data cover the financial crisis of $2008$-$2009$. For related discussion, yet focused on principal components as opposed to nonlinear macros, see \cite{DTKK2021b}. In what follows, we only concentrate on models $GP_{110}(CPI)$ and $GP_{011}(GRO)$, which we also focus on in Section \ref{sec:RPFandHiddenNonlin}, for specific reasons mentioned therein. %Favourable outcome of the tuning procedure, as outlined in Appendix \ref{appendix:GPcalibration}, makes these models also well suited for the purposes of such a visual investigation.      

By inspecting the %scatter 
plots for model $GP_{110}(CPI)$, which are shown in Figure \ref{fig:GP110CPIscatters}, we observe that $CPI$ interacts with the yield curve in a nonlinear, rather than linear, manner. This is particularly the case when we look at panels on the right hand side of this figure. In there, distributions of $v^{(1)}(CPI_{t-1})$ and $v^{(2)}(CPI_{t-1})$ plotted against $CPI_{t-1}$ are estimated using the entire sample of data, including the recession period. Comparing these two panels with their counterparts on the left hand side in the same figure, which are based on data from the training period only, and thus preclude the recession, we notice that the nonlinear relationship between $CPI$ and the yield curve strengthens in time. %This observation is in line with the message coming from $\log(\ell_1^{-1})$ and $\log(\ell_2^{-1})$ in Figure \ref{fig:GP111CPIibis}, where these log relevance scores initially increase in the testing period and then, approximately half way through, stabilize in time.  

%Similar observations are to be made in Figure \ref{fig:GP111F8ibis}, which illustrates output from model $GP_{111}(F8)$. There, for $F8$ we observe analogous dynamics of the corresponding log relevance score as it is for $CPI$. However, when inspecting the related posterior means of $f^{(j)}(F8_{t-1})$, $j=1,2,3$, from model $GP_{111}(F8)$ in Figure \ref{fig:FsF8} therein, and also those from models $GP_{011}(F8)$ and $GP_{001}(F8)$, it is rather a nontrivial task to make observations as evident as these in case of $CPI$ in Figure \ref{fig:FsCPI}.Therefore, we instead 
Next, we have a careful look at $GRO$, where things are quite different. In Figure \ref{fig:GP011GROscatters}, with associated plots for model $GP_{011}(GRO)$, we observe that $GRO$ interacts with the yield curve more in a linear than in a nonlinear manner. It is particularly evident when inspecting the top panel in this figure, which shows the distribution of $v^{(2)}(GRO_{t-1})$ plotted against $GRO_{t-1}$. %The case of $v^{(1)}(GRO_{t-1})$ remains rather inconclusive. 
The vast majority of data points in the training period, that is on the left hand side, are located in the part of the graph where the association is linear. The remaining data points constitute the outliers. On the right hand side, that is when the entire data sample is used, % the testing period, 
the part of the graph which is linear is accompanied by even more outliers. With hindsight, economic activity is heavily affected by outliers due to recession, what can also be %It becomes
noticed by comparing %the ranges on 
the horizontal axes in these plots between the left and the right hand side panels. %Because of the outlying observations,
Although the functional association of lagged $GRO$ with the yield curve %, particularly this 
shown in the top %and mid 
panel deteriorates only slightly in time, %. However
this presented in the bottom panel for $v^{(3)}(GRO_{t-1})$ %which is difficult to identify already in the training period, 
dilutes completely when the entire data sample is involved. Likely, the latter changes in the post-recession period, while the model assumes it is the same, what makes the forecasting exercise we conduct quite challenging but at the same time more realistic.

\subsection{Bond Return Predictability}% and Economic Performance}

%In this section we present and discuss the results on bond return predictability and the associated %economic performance of our nonlinear, and also linear, models, as it is showcased in Section %\ref{sec:predEconomicValue}. Where insightful, we also link these outcomes to the %in-sample 
%explanatory power behind the observed (linear) and the estimated (nonlinear) macros, where selected %results are shown in Tables \ref{table:deltaR2adj0} and \ref{table:deltaR2adj111}. %%\ref{table:deltaR2adj001}).

%\subsubsection{Predictive Performance}

In what follows, we focus on statistical performance. Table \ref{table:R2OS} reports out-of sample $R^2$ values for all linear (including $M_0$, the maximally flexible model, and $M_1$, see \cite{DTKK2021b}) and nonlinear models across different bond maturities and at 1-month prediction horizon. The latter means that we only consider non-overlapping excess bond returns. The in-sample (training) period is from January 1985 to the end of 2007 and the out-of-sample (testing) period is from January 2008 to the end of 2018. The latter practically begins with the $2008$-$2009$ financial crisis. Overall, results indicate, as expected and in line with the existing literature \citep{Duffee11,Joslin14,Fulop19}, that models which incorporate macroeconomic information, irrespective of whether in a linear or nonlinear manner, predict well and perform better out-of-sample compared to the EH benchmark, as well as when set against yields-only models. This is confirmed by predictive $R^2_{os}$ that are exclusively positive and statistically significant for $CPI$, across all maturities, and mostly positive and statistically significant for $GRO$, especially for the short and medium term maturities. Across maturities, they range from $0.02$ to $0.06$ for $CPI$, whereas for $GRO$ the positive and statistically significant $R^2_{os}$ are between $0.02$ to $0.05$. 

Generally, models with macroeconomic variables perform better, in terms of out-of-sample statistical performance, than the yields-only models we consider, namely $M_0$ and $M_1$. However, there are four notable exceptions in case of $GRO$, where $R^2_{os}$ are practically close to zero. These are linear and nonlinear models with indices $111$ and $110$, which either perform as unsatisfactorily as $M_1$ ($GP_{111}$) or worse, and as such similar to $M_0$ ($LM_{111}$, $LM_{110}$ and $GP_{110}$). What they interestingly have in common is that, in these models macroeconomic information interacts with the yield curve through the observed level, meaning that it is included in the equation governing the real-world dynamics of the first PC. This deterioration in statistical performance translates further into correspondingly poor economic value results, especially for linear models, what we discuss in the next section.

Selected results on %in-sample 
explanatory power for the nonlinear models with $GRO$ specifically listed above, which %are obtained using data ranging from the beginning of 1985 until the end of 2017, and 
are presented in Table \ref{table:deltaR2adj111} (see fourth panel) for model $GP_{111}$, %and \ref{table:deltaR2adj110} (see third panel), 
reveal that, contrary to the out-of-sample statistical performance results, there are statistically significant benefits from including $v^{(1)}(GRO_{t-1})$, when explaining the variability of excess bond returns. It is particularly evident when compared to what we observe for the corresponding nonlinear function of lagged $CPI$. For $GRO$, such gains decrease at longer maturities and range from $2.91\%$ at 2-year, through $2.17\%$ at 5-year, to $0.34\%$ at 10-year maturity. For $CPI$, these gains are considerably smaller and amount to about $0.60\%$ throughout 2- to 5-year maturities. Related results in Table \ref{table:deltaR2adj0} confirm that similar is true in the linear case. Namely, explanatory power gains for $GRO$, albeit smaller than in the nonlinear case, range from $1.95\%$ at 2-year to $0.56\%$ at 4-year maturity, yet occur only at the short end of the yield curve. Those for $CPI$ are in this case practically negligible. Corresponding results on explanatory power for model $GP_{110}$ are quantitatively and qualitatively alike, thus not shown. The latter also applies to such results for all models with indices $011$, $010$ and $001$.  

\begin{figure}[t]
\begin{center}
\includegraphics[trim = 20mm 0mm 0mm 0mm, height=5.2in,left]{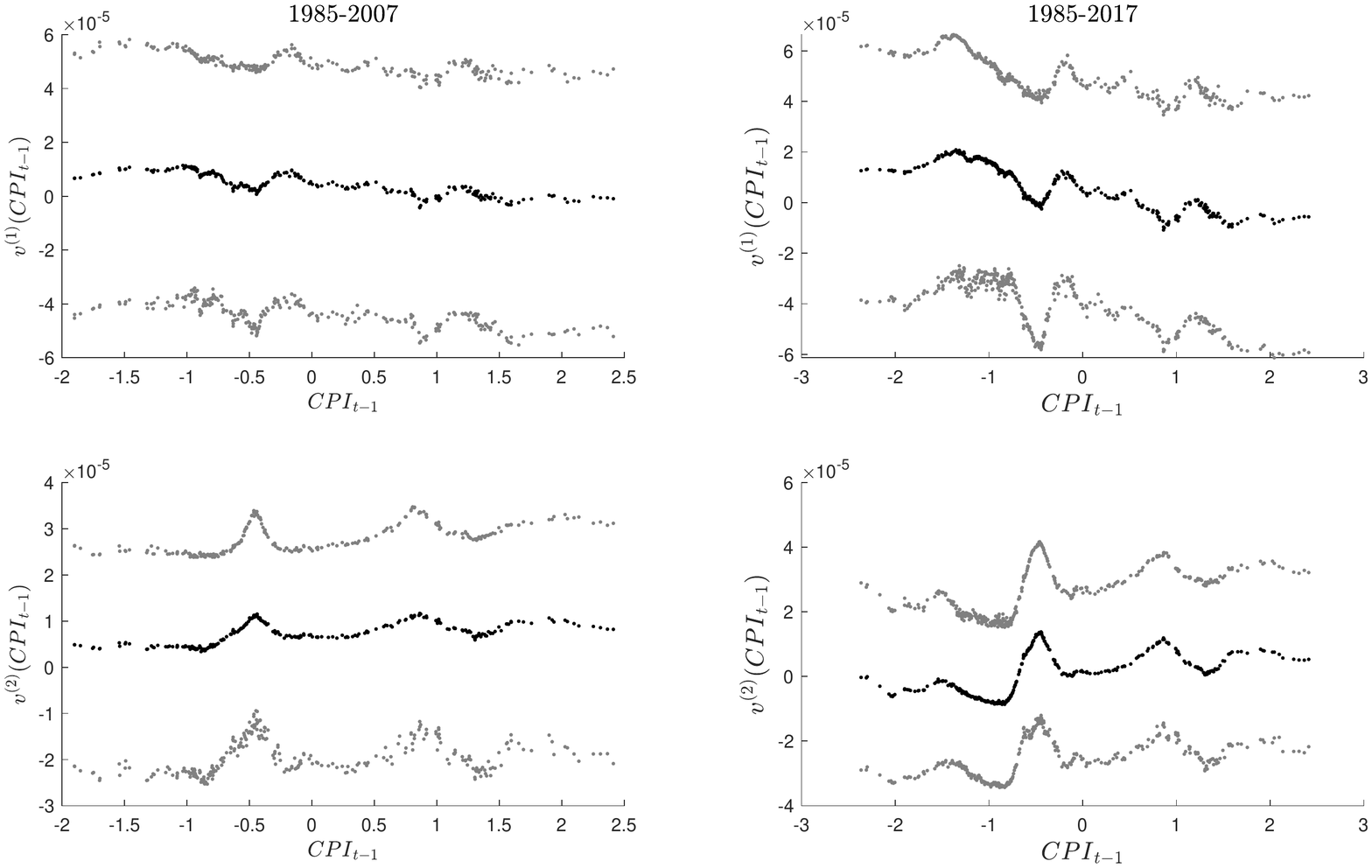}
\caption[Plots of lagged $CPI$ against its nonlinear function $v$ from model $GP_{110}$]{Plots of lagged $CPI$ against its nonlinear function $v$, obtained from model $GP_{110}$. Plots in the left column are based on model parameters estimated using data from the training period only (January $1985$ - end of $2007$). Plots in the right column result from model parameters estimated using the entire sample of data (January $1985$ - end of $2017$). Throughout, points in black correspond to posterior mean of $v$ and those in grey to its 95\% credible intervals, all calculated from the IBIS output.} 
\label{fig:GP110CPIscatters}
\end{center}
\end{figure}

\begin{figure}[t]
\begin{center}
\includegraphics[trim = 20mm 0mm 0mm 0mm, height=5.2in,left]{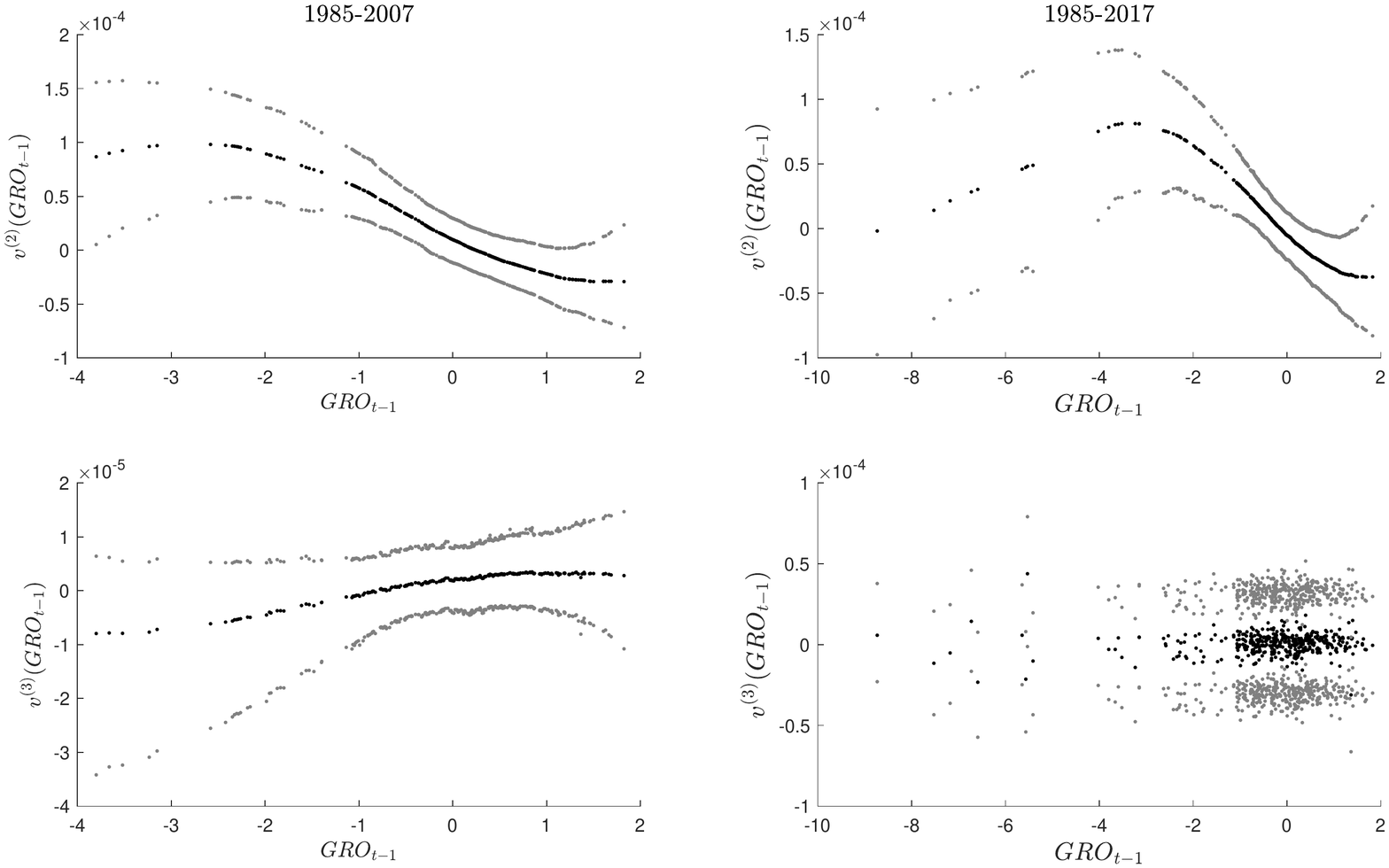}
\caption[Plots of lagged $GRO$ against its nonlinear function $v$ from model $GP_{011}$]{Plots of lagged $GRO$ against its nonlinear function $v$, obtained from model $GP_{011}$. Plots in the left column are based on model parameters estimated using data from the training period only (January $1985$ - end of $2007$). Plots in the right column result from model parameters estimated using the entire sample of data (January $1985$ - end of $2017$). Throughout, points in black correspond to posterior mean of $v$ and those in grey to its 95\% credible intervals, all calculated from the IBIS output.} 
\label{fig:GP011GROscatters}
\end{center}
\end{figure}

One possible rationale behind the above argument is that $GRO$, the economic activity, varies more profoundly than $CPI$, the prices, in the aftermath of the $2008$-$2009$ financial crisis, that is throughout the testing sample. Let us compare the horizontal axes on the left with those on the right hand side in the plots shown in Figures \ref{fig:GP110CPIscatters} and \ref{fig:GP011GROscatters}. In Figure \ref{fig:GP110CPIscatters}, $CPI_{t-1}$ is approximately between $-2$ and $2.5$ over 1985-2007 and from $-2.5$ to $2.5$ over 1985-2017. In Figure \ref{fig:GP011GROscatters}, $GRO_{t-1}$ is approximately between $-4$ and $2$ over 1985-2007 and from $-10$ to $2$ over 1985-2017. Importantly, both variables are standardized over 1985-2017 based on their respective means and standard deviations computed over 1985-2007. Thus, throughout 1985-2017 the variation of $CPI$ hardly moves relative to what happens for $GRO$, which becomes evidently negatively skewed as result of such adverse economic conditions. On the other hand, these abnormal changes in $GRO$ serve explaining the variability of excess bond returns in these turbulent times well, what is also a sign of possible overfitting.
%See also Tables \ref{table:deltaR2adj011}, \ref{table:deltaR2adj010} and \ref{table:deltaR2adj001} where related results 

%Finally, it is not without reason that above argumentation concerns macroeconomic impact on the yield curve level, which is of highest magnitude out of all PCs. Failure to predict it accurately affects excess return predictions most. This is because associated errors stemming from incorporating macros translate more deeply into mistakes made when forecasting excess returns, especially if set against such errors which only impact slope and curvature of the yield curve. A preview comparison of results in Table \ref{table:CER} (see second panel) for models with indices 111 and 110, with those for models with indices 011 and 001, further facilitates this argument, for both linear and nonlinear cases, in terms of economic value. 
Nevertheless, if the underlying index does begin with 1 not 0 then these nonlinear models with $GRO$ we specifically mention above, with indices $111$ and $110$, tend to significantly improve out-of-sample statistical performance results when compared to linear benchmarks, see Table \ref{table:R2OSLM}. As such, they reduce the risk associated with out-of-sample predictions based on macroeconomic variables which become negatively skewed in times of economic/financial crises. Interestingly, similar improvements occur for $CPI$ in the specific case of nonlinear model with index $110$. %Corresponding results for $CPI$ are overall not convincing, except for the case of $110$.

\subsection{Economic Performance}%Value}

In what comes next, we concentrate on model performance in terms of economic value. Table \ref{table:CER} reports results for the annualized $CER$s, calculated using out-of-sample forecasts of bond excess returns across maturities and at 1-month prediction horizon. %As such, bond returns herein are non-overlapping. 
The in-sample and out-of-sample periods are the same as in the case of predictive performance. 
The coefficient of relative risk aversion is chosen to be $\gamma = 3$ and we do not impose any portfolio constraints\footnote{These non-conservative choices are motivated by early exploratory character of the analysis conducted herein. The goal is to find economic value first and in future research examine if it can be exploited when stricter conditions apply.}. We find that, in most cases, corresponding $CER$s are positive and non-negligible, indicating that DTSMs with unspanned macroeconomic information, introduced in the model in a linear or nonlinear way, not only perform well and in a statistically significant manner when it comes to out-of-sample predictability but also generate economic gains for bond investors relative to the EH benchmark.

For the linear case, this is in line with the existing literature, as numerous studies show \citep{Duffee11,Joslin14,Fulop19}, whereas in case of nonlinear models only recently \cite{Bianchi20}, albeit in a regression and not in a no-arbitrage setting, show that forecasts based on neural networks fed with macroeconomic and yield information jointly, translate into economic gains that are larger than those obtained using yields alone. In line with all that, yields-only DTSMs we are concerned with in this paper, that is $M_0$ and $M_1$, perform evidently worse in terms of economic value than models with macroeconomic information (see, Table \ref{table:CER}). This is also consistent with their relatively inferior out-of-sample statistical performance, which we discuss in the previous section.

In case of $CPI$ results show clear evidence of positive out-of-sample economic benefits for bond investors from introducing information about prices in the model. This is the case across most maturities (especially at 2-, 3- and 7-year) and in particular when $CPI$ is introduced in a nonlinear manner. $CER$s for linear models with $CPI$ are not even comparably as statistically significant as the former. For example, in case of $GP_{010}$ they are statistically significant at all maturities and range from $3.67\%$ (at 5-year) to $4.94\%$ (at 7-year).  
In relation to $GRO$ such evidence is less pronounced and results are only statistically significant for selected maturities and for linear and nonlinear models with indices 011 and 010 (at 3-, 4- and 7-year, in both cases), as well as 001 (all maturities but 5-year). The outcome is most positive in this last case and, when statistically significant, $CER$s for linear and nonlinear models with index 001 range from $3.37\%$ (at 3-year, for the linear setup) to $4.55\%$ (at 7-year, for the nonlinear setup).

When comparing nonlinear models with their linear counterparts, only in case of $CPI$ the former perform significantly better than the latter (see, Table \ref{table:CERLM}). For $GP_{110}$, $GP_{010}$ and $GP_{001}$ we observe positive and non-negligible $CER$ values across most maturities, in particular at 2- to 5-year. If statistically significant, they span between $1.88\%$ and $3.78\%$ relative to the linear benchmarks. For $GRO$, where results are overall worse compared to those for $CPI$, we are not able to identify meaningful, statistically significant cases where nonlinear models outperform their linear benchmarks in terms of economic value. In the next section we provide some rationale why and when it might be so. 

Overall, out-of-sample economic value results are consistent with those for predictive performance which we discuss in the previous section. As such, based on results presented in this paper, the puzzling behavior between statistical predictability and out-of-sample economic gains for bond investors cannot be unequivocally confirmed for DTSMs utilizing macroeconomic information, especially in a nonlinear manner. %This also includes those arguments which reveal themselves from the analysis of explanatory power gains in-sample, as presented in Tables \ref{table:deltaR2adj0}-\ref{table:deltaR2adj001}. Namely, models with $CPI$, both linear and nonlinear, lead to superior economic value results out-of-sample compared to such models with $GRO$ and the same happens in case of their predictive performance. 
It is also important to note that considerable and statistically significant explanatory power gains (see, Tables \ref{table:deltaR2adj0} and \ref{table:deltaR2adj111}), especially from macroeconomic variables which are prone to outliers like $GRO$, not necessarily translate into corresponding economic performance out-of-sample. Even to the contrary, they may lead to significant losses, especially when economic conditions are changing. It is particularly the case for linear models with indices $111$ and $110$ (see, Table \ref{table:CER}). %Situation is similar with the associated predictive performance. 
However, nonlinear models may help limit such losses, e.g. by handling outliers better. In some cases, like for $CPI$ ($GP_{110}$, $GP_{010}$ and $GP_{001}$), they perform consistently better out-of-sample than their linear counterparts in terms of economic value (see, Table \ref{table:CERLM}).

\subsection{Benefiting from Hidden Nonlinearities}
\label{sec:RPFandHiddenNonlin}

In this section we shed some light on why and when certain nonlinear models we consider in this paper (with macroeconomic variables like $CPI$) outperform their linear counterparts in terms of economic value, whereas other (with macros such as $GRO$) do not. To that end, we resort to the results obtained from risk premium factor decomposition, as described in Section \ref{subsec:modelsrationale}, which are presented in Tables \ref{table:R2adjRPFEFPFM001010110} and \ref{table:R2adjRPFEFPFM011111}. For pragmatic reasons, related to the tuning procedure  (see, Appendix \ref{appendix:GPcalibration}) which is not always optimal and as such affects the way certain aspects of this decomposition can be effectively demonstrated graphically, in what follows we focus on two particular models. Namely, $GP_{110}(CPI)$ and $GP_{011}(GRO)$.

%However, before doing that we take a closer look at results in Table \ref{table:R2adjFP}, where we can inspect how well the PCs from observed yields explain nonlinear macros $v^{(j)}(M_{t-1})$, $j\in\{1,2,3\}$, for $CPI$ and $GRO$, across models. It is clear and not surprising that for $CPI$ the adjusted $R^2$, our metric of choice for linear association, are much higher than they are for the $GRO$. In case of $v^{(1)}(M_{t-1})$, they range from $0.64$ to $0.68$ for the former and are between $0.04$ and $0.05$ for the latter. For $v^{(2)}(M_{t-1})$, respective $\bar{R}^2$ are between $0.29$ and $0.53$ and from $0.07$ and $0.09$. When it comes to $v^{(3)}(M_{t-1})$, these are in ranges of $0.05$-$0.54$ and $0.02$-$0.10$, for $CPI$ and $GRO$, respectively. Differences for given nonlinear macro across models stem from varying tuning outcomes. %calibrations, as shown in Table \ref{table:SIGs}. 
%These results are consistent with the stylised fact that information about the prices is in the yield curve, wheres this on economic activity isn't \citep{Duffee11,Joslin14}. This is also confirmed by the outcomes we present in Table \ref{table:deltaR2adj0}, where adding lagged $CPI$ to a regression explaining excess bond returns on top of all three PCs has no incremental value, while doing the same with lagged $GRO$ (statistically) significantly adds explanatory power to this regression, at least for some maturities (2Y, 3Y and 4Y). 

We choose these two models for several reasons. First, both linear and nonlinear models with $CPI$ and index 110 perform well in terms of out-of-sample predictability, however the nonlinear model performs significantly better across maturities than its linear counterpart - with $R^2_{os}$ ranging from $0.04$ to $0.06$ against $0.02$ and $0.03$ for the linear model (see, Table \ref{table:R2OS}). Second, both linear and nonlinear models with $GRO$ and index 011 perform fairly well across most maturities when it comes to out-of-sample predictability - with statistically significant $R^2_{os}$ ranging from $0.02$ and $0.05$ for both cases. Third, while nonlinear model with $CPI$ does significantly better in terms of out-of-sample economic performance than its linear counterpart - with relative $CER$ ranging from $2.55\%$ to $3.35\%$ across maturities when statistically significant - it is almost entirely the opposite case for the $GRO$. There, relative $CER$ are not statistically significant and vary between $-3.05\%$ and $1.59\%$ (see, Table \ref{table:CERLM}). This is precisely what we are after to facilitate the argumentation which follows.

With what we summarised above in mind, we now resort to decomposition in \eqref{RPFreg} of nonlinear macro $v(M_{t-1})$ into a part about which complete information is contained in the yield curve, that is $E[v(M_{t-1})|\mathcal{P}_t]$, and a part which is entirely hidden, that is $\widetilde{RP}^V_t(M_{t-1})$. We regress individually each of these components on the underlying macro $M_{t-1}$ in order to determine the degree to which the resulting relationship is linear, as measured by the associated adjusted $R^2$. Tables \ref{table:R2adjRPFEFPFM001010110} and \ref{table:R2adjRPFEFPFM011111} contain such results for $GP_{110}(CPI)$ and $GP_{011}(GRO)$, respectively. They include related results for all the other models as well, however our focus on these two remains unchanged. 

\begin{figure}[t]
\begin{center}
\includegraphics[trim = 20mm 0mm 0mm 0mm, height=5.2in,left]{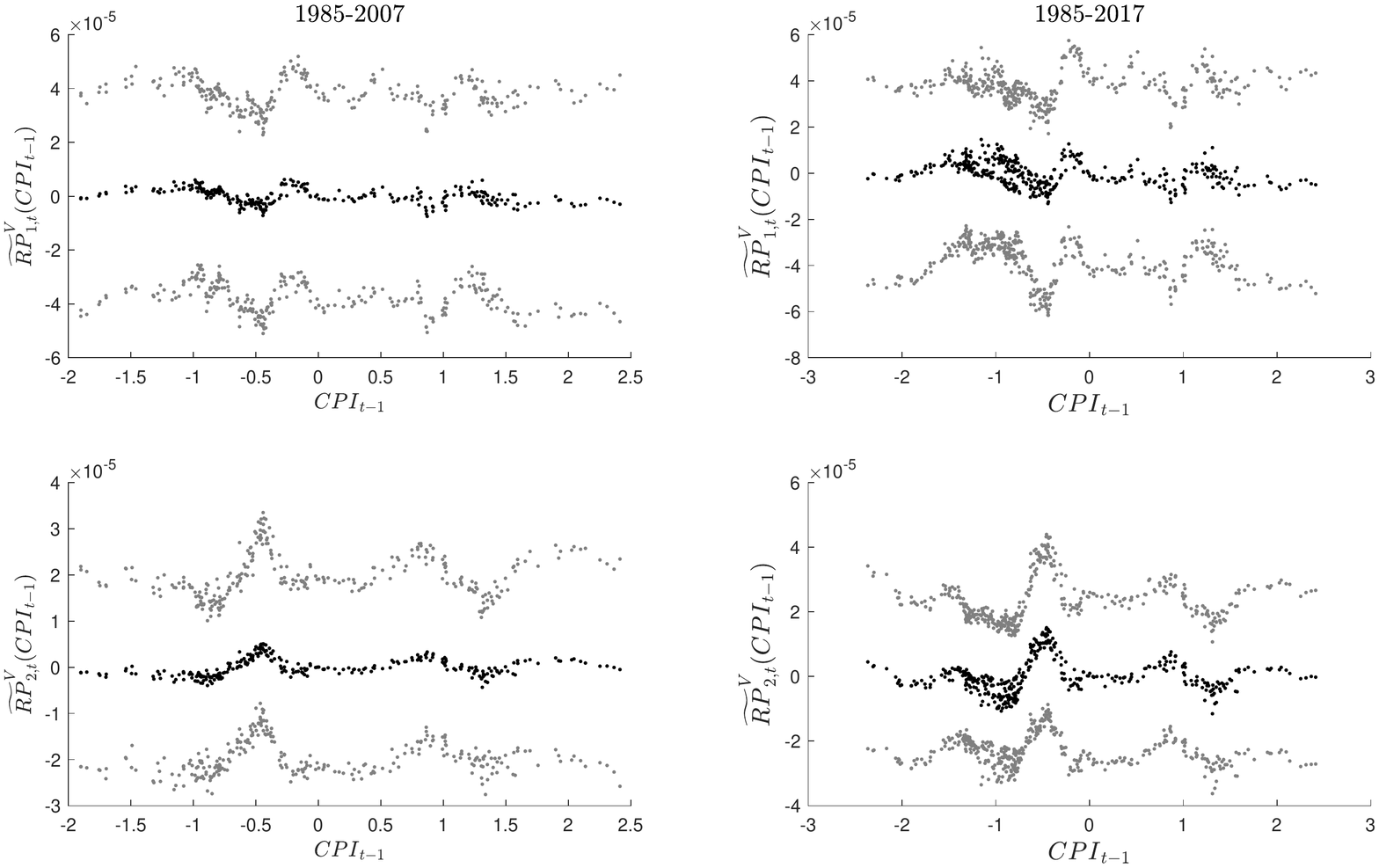}
\caption[Plots of lagged $CPI$ against the hidden part $\widetilde{RP}^V$ of its nonlinear function $v$ from model $GP_{110}$]{Plots of lagged $CPI$ against the hidden part $\widetilde{RP}^V$ of its nonlinear function $v$, obtained from model $GP_{110}$. Plots in the left column are based on model parameters estimated using data from the training period only (January $1985$ - end of $2007$). Plots in the right column result from model parameters estimated using the entire sample of data (January $1985$ - end of $2017$). Throughout, points in black correspond to posterior mean of $\widetilde{RP}^V$ and those in grey to its 95\% credible intervals, all calculated from the IBIS output.} 
\label{fig:GP110CPIscattersRPF}
\end{center}
\end{figure}

Beginning with $CPI$, in the third panel of Table \ref{table:R2adjRPFEFPFM001010110} we notice that for model $GP_{110}(CPI)$ both $v^{(1)}(CPI_{t-1})$ and its part which is not hidden from the yield curve are to large extent linear functions of $CPI_{t-1}$, as measured by $\bar{R}^2$ that are equal to $0.73$ and $0.70$, respectively. At the same time the hidden part, that is $\widetilde{RP}^V_{1,t}(CPI_{t-1})$, is hardly in a linear relationship with $CPI_{t-1}$, as indicated by a very low value of $\bar{R}^2$ equal to $0.09$ only. It is similar in case of $v^{(2)}(CPI_{t-1})$ where these metrics amount to $0.32$, $0.67$ and practically $0.00$, in that order. Hidden parts of $v(CPI_{t-1})$ in model $GP_{110}(CPI)$ are thus nonlinear functions of the underlying macroeconomic variable, namely $CPI_{t-1}$. To realize that it is indeed so, it is sufficient to have a brief look at Figure \ref{fig:GP110CPIscattersRPF}. Therein, we see clearly that functional associations between these hidden parts and lagged $CPI$ are highly nonlinear, stable or become even more pronounced (left versus right hand side). %, where the latter fact is further supported by the dynamics in time of the underlying log relevance scores in the top right panel of Figure \ref{fig:GP110CPIibis}.
At this point it is safe to state that in this particular case there is a fair reason behind superiority of the nonlinear model, which is implemented using $GP$s in this paper, over its linear counterpart. Namely, unspanned information coming from $CPI$, which is hidden from the yield curve yet it affects its $\mathbb{P}$-dynamics, is evidently nonlinear in nature.

Ending with $GRO$, in the first panel of Table \ref{table:R2adjRPFEFPFM011111} we observe that, for model $GP_{011}(GRO)$, both $v^{(2)}(GRO_{t-1})$ and its part which is not hidden from the yield curve are to considerable and to moderate extent, respectively, linear functions of $GRO_{t-1}$, as measured by $\bar{R}^2$ that are equal to $0.50$ and $0.17$. What is however different from the case of $CPI$ is that the hidden part, namely $\widetilde{RP}^V_{2,t}(GRO_{t-1})$, is to larger extent linear in $GRO_{t-1}$ than the part which is not hidden from the yield curve, that is $E[v^{(2)}(GRO_{t-1})|\mathcal{P}_t]$, as indicated by $\bar{R}^2$ equal to $0.37$ for the former. In case of $v^{(3)}(GRO_{t-1})$, these metrics amount to $0.00$, $0.15$ and $0.03$, in that order. However, for reasons mentioned in the last paragraph of Section \ref{subsec:YCandNonlinDyn} and in relation to the tuning procedure (see, Appendix \ref{appendix:GPcalibration}), which is not always optimal,  %this is due to the caveats of calibration procedure we devise (see, Appendix \ref{appendix:GPcalibration}), which are discussed in greater detail earlier in Section \ref{subsec:YCandNonlinDyn}, 
we also look at the corresponding numbers for model $GP_{001}(GRO_{t-1})$ in the first panel of Table \ref{table:R2adjRPFEFPFM001010110}. They amount to $0.39$, $0.11$ and $0.33$, respectively. In this case, the focus of the tuning procedure lies solely on the curvature not slope and curvature together, as in the former model. These latter results correspond more closely to what we observe for $v^{(2)}(GRO_{t-1})$ in model $GP_{011}(GRO_{t-1})$.

\begin{figure}[t]
\begin{center}
\includegraphics[trim = 20mm 0mm 0mm 0mm, height=5.2in,left]{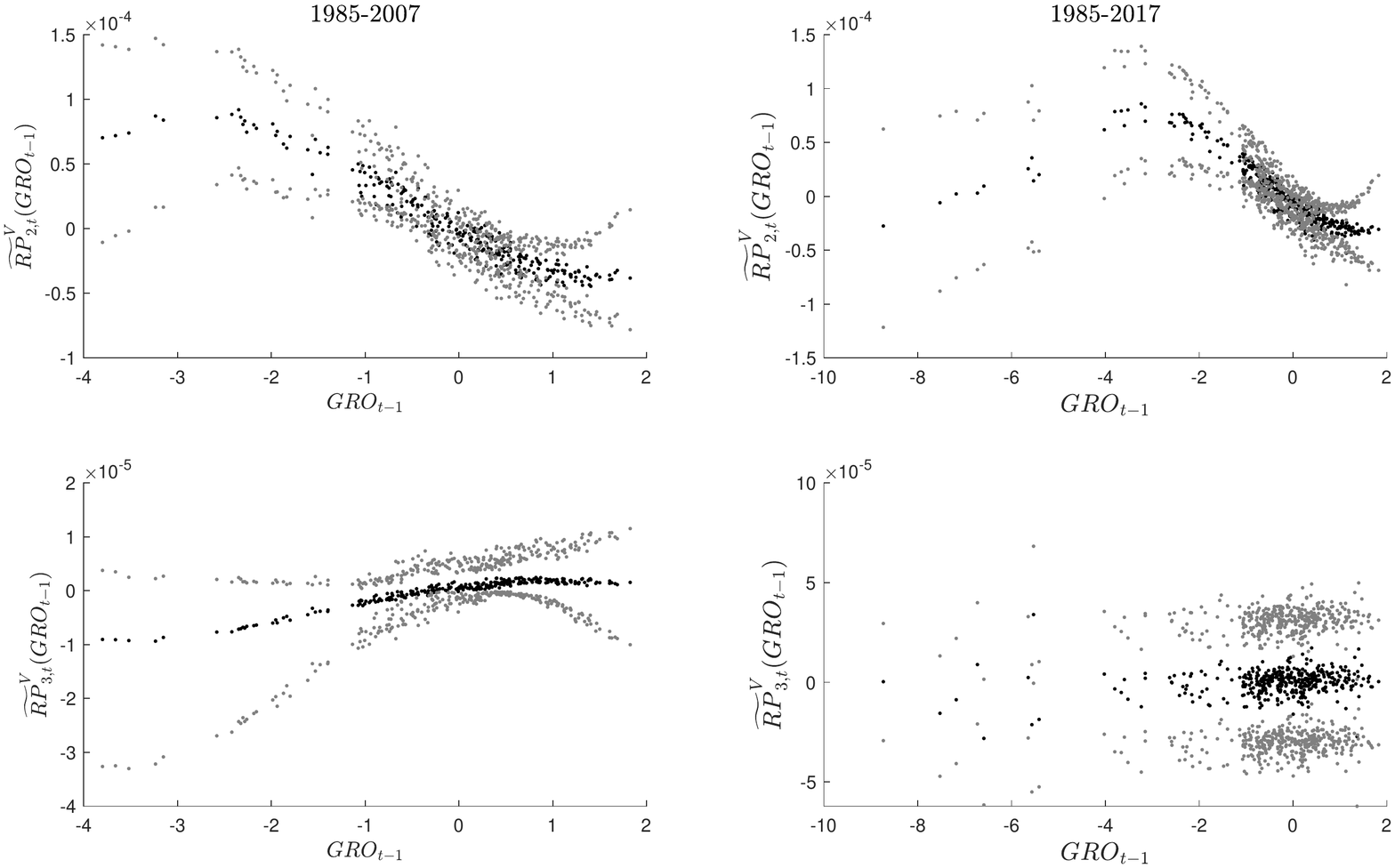}
\caption[Plots of lagged $GRO$ against the hidden part $\widetilde{RP}^V$ of its nonlinear function $v$ from model $GP_{011}$]{Plots of lagged $GRO$ against the hidden part $\widetilde{RP}^V$ of its nonlinear function $v$, obtained from model $GP_{011}$. Plots in the left column are based on model parameters estimated using data from the training period only (January $1985$ - end of $2007$). Plots in the right column result from model parameters estimated using the entire sample of data (January $1985$ - end of $2017$). Throughout, points in black correspond to posterior mean of $\widetilde{RP}^V$ and those in grey to its 95\% credible intervals, all calculated from the IBIS output.} 
\label{fig:GP011GROscattersRPF}
\end{center}
\end{figure}

To consolidate the discussion, a careful look at Figure \ref{fig:GP011GROscattersRPF} is required. In the top panel, we notice immediately that what adjusted $R^2$ are telling us about $v^{(2)}(GRO_{t-1})$ in the model of interest is indeed so. Namely, the association between its hidden part and the lagged $GRO$ is visibly more linear than nonlinear. The picture is similar, albeit less pronounced, to this in the corresponding panel in Figure \ref{fig:GP011GROscatters}, which we discuss in the last paragraph of Section \ref{subsec:YCandNonlinDyn}. On both sides the majority
of data points are located in the part of the graph where the association is linear, whereas the remaining data
points are the outliers.   
%, although there are some considerable outliers, and importantly stable (left versus right hand side). %, where the latter finding is clearly supported by the dynamics in time of the associated log relevance score in the top right panel of Figure \ref{fig:GP011GROibis}. 
What happens to $\widetilde{RP}^V_{3,t}(GRO_{t-1})$ in the bottom panel can also be linked to related arguments in the same section. %the caveats of the calibration procedure we mention above. On the left hand side in the bottom panel of this figure, we see that at least over the period 1985-2007 the relationship between the hidden part of $v^{(3)}(GRO_{t-1})$ and lagged $GRO$ has definitely linear character. On the right hand side, we witness again the effects of imperfect model calibration, which blur the true nature of the functional association between $\widetilde{RP}^V_{3,t}(GRO_{t-1})$ and $GRO_{t-1}$. Nevertheless, we are convinced that 
It is thus not surprising that in this case the nonlinear model leads to inferior economic value results, let alone adds such a value, in comparison to the linear benchmark (see, Table \ref{table:CERLM}). In this case and on such front it is evidently hard to beat the linear, more parsimonious model based merely on outliers. %For $GRO$, the unspanned information is hidden from the yield curve, but it affects its $\mathbb{P}$-dynamics, is rather linear in nature than it isn't.

\section{Conclusions}
\label{sec:Conclude}

We propose a novel methodological framework which combines Bayesian sequential inference with machine learning techniques, in particular Gaussian Processes. It allows us to incorporate unspanned macroeconomic information into Dynamic Term Structure Models in a potentially nonlinear manner. Sequential setup we develop successfully handles real-time adjustments to parameters governing such asymmetric/nonlinear associations. %It is also capable of monitoring their stability in time. 
The methodology takes into account parameter uncertainty and provides entire predictive distribution of bond returns, allowing investors to review their beliefs when new information arrives and thus informing their asset allocation in an online manner. The framework is then tested against the Expectations Hypothesis, as well as against the linear benchmark, in a comprehensive out-of-sample exercise involving statistical predictability and economic value, where we assume availability of information about prices and economic activity. To that end, we scrutinize nonlinear models by developing a version of the macro-finance DTSM proposed by \cite{Joslin14} that incorporates unspanned macroeconomic variables in a linear manner. However, to align with our nonlinear setup we consider them exogenous. 

Empirical results confirm that in such an exercise the models with unspanned macroeconomic information, irrespective if it is introduced in a linear or nonlinear manner, perform overall better than the models which are yields-only by construction. Specifically, they also reveal that nonlinear models provide a competitive edge over linear models in cases when \textit{a-priori} possibly nonlinear functions of macroeconomic variables admit nonlinear character \textit{a-posteriori} in the parts which are hidden from the yield curve. In this paper it is the case for the prices and not for the economic activity. To demonstrate that, we apply the risk premium factor decomposition from \cite{Duffee11}.

However, results in this paper also indicate that for certain models where macroeconomic information directly affects the yield curve level, using nonlinear formulation might not necessarily lead to superb predictability and economic value gains but it might still mitigate substantial mistakes on those fronts. In particular, it is the case for models with macroeconomic variables which in abnormal times, such as the $2008$-$2009$ financial crisis or the more recent COVID-$19$ recession, are subject to outliers like the economic activity. According to our results, Gaussian Processes handle such cases quite well.

Thinking ahead, we acknowledge the limitations of our study where we only consider a set of macroeconomic variables limited to two and Gaussian Processes we apply are of a single-input type and based on one commonly applied nonlinear kernel that we select to use. Verifying what we infer in this paper on a broader set of macroeconomic information, while also using other nonlinear kernels such as Matern kernel or combined kernels, is a possible way going forward. Letting macroeconomic variables interact within a multi-input Gaussian Process framework, for example using ARD (Automatic Relevance Determination) kernels \citep{RasmussenWilliams2006}, is also a potentially interesting %andstraightforward 
research avenue to pursue. %Another intriguing, future direction is to incorporate spanned instead of unspanned nonlinear macros %, what developed framework can be adopted to, 
%and embark on policy oriented applications, what such an extension facilitates.

\bibliography{referencefile3_ARXIV}

\begin{appendices}

\section{Specification of Priors}
\label{appendix:priors}

In what comes next, we succinctly explain %provide
the prior distributions that were not specified in the main body of the paper. For parameters in $\Sigma_{\mathcal{P}}$, $g^{\mathbb{Q}}$, $k_{\infty}^{\mathbb{Q}}$, $\sigma_e^2$ and $\lambda^{\mathcal{P}}$, or effectively $\lambda_{1,2}$, priors are constructed in the same manner as in the related appendix in \cite{DTKK2021a}. 

The only exception are parameters in $\ell_K$, which have range restricted to positive values.
%Low informative priors on each $\theta$ component can be assigned although some relevant information is available in this context \citep{Chib09}.
We thus transform them first, %restricted range
%parameters
so that they have unrestricted range. %We also scale parameters which typically take very small values. 
Specifically, we work in log-scale of $\ell_K$. % and consider a Cholesky factorization of $\Sigma_{\mathcal{P}}$ where the diagonal elements are transformed to the real line and off-diagonal elements are scaled by $10^4$. In order to preserve the ordering of the eigenvalues $g^{\mathbb{Q}}$ we apply a reparametrization and work with their increments that are again transformed to the real line. Finally, we scale $k_{\infty}^{\mathbb{Q}}$ by $10^6$.
%(\textbf{TDT: We do this first: diagonal = diag(K1QX); then this: K1QX=diag(-exp(-[diagonal(1);diff(diagonal)])); To transform back we do: diagonal=diag(K1QX);and this: diagonal=-log(-diagonal); plus this: K1QX=diag(cumsum(diagonal)); I am not sure it exactly means that we just transform to the real line eventually, there are increments but we exponentiate them actually on top of this. This is actually how we do it in the MCMC, in the MLE step we use logit to achieve ordering. I am not sure how to describe it clearly though. This exponentiation I recall improved acceptance rates, proposal got better, more robust.}) 
Next, independent normal distributions with zero means and large variances are assigned to each of its components.%$\theta$, except $\lambda^{\mathcal{P}}$, as explained below, and $\sigma_e^2$, where we assign a conjugate Inverse-Gamma prior as in \cite{Bauer18} with parameters $\alpha/2$ and $\beta/2$. We take $\alpha=\beta=0$ for an entirely diffuse case.     
%In general, prior for $\lambda^{\mathcal{P}}$ cannot be uninformative because it would cause Bayes factors indeterminate \citep{KassRaftery1995}. Although immaterial in large samples, its choice plays a significant role for inference in small samples that one is dealing with in DTSM setting. Following \cite{Bauer18}, prior specification for $\lambda_{1,2}$, which is the only unrestricted risk price in the model, involves a normal distribution centered around zero with some variance. To estimate this variance, similarly to the case when more risk prices were unrestricted, we assume conditional prior independence between elements of $\lambda^{\mathcal{P}}$. To that end, we use orthogonalized $g$-prior, where $g$ is equal to the number of observations. Specifically, the covariance matrix of such a $g$-prior distribution, which is a normal, is proportional to that of the least-squares estimator but with off-diagonal elements equal to zeros. To obtain it, all standard case model parameters other than $\lambda^{\mathcal{P}}$ are set to their maximum likelihood estimates and all risk prices are assumed unrestricted. Hence, we are specifying a yields-only, maximally flexible DTSM. Then, least-squares estimates of $\lambda^{\mathcal{P}}$ are calculated. Eventually, we have that $g$ times the diagonal element of the covariance matrix for the resulting least-squares estimates of risk prices which corresponds to $\lambda_{1,2}$ is the variance we sought.

\section{Markov Chain Monte Carlo Scheme}
\label{appendix:MCMC}

Following from \eqref{likelihoodlambda12} and \eqref{posterior}, and given a prior $\pi(\theta)$ as described in Appendix \ref{appendix:priors}, the posterior can be written in a more detailed manner as
\begin{equation}\label{posterior2}
\begin{array}{l}
\pi(\theta|Y,M,\widehat{\sigma}_K) = \left\{\prod_{t=0}^T f^{\mathbb{Q}}(y_{t}|\mathcal{P}_{t}, k_{\infty}^{\mathbb{Q}}, g^{\mathbb{Q}}, \Sigma_{\mathcal{P}}, \sigma_{e}^{2})\right\} \times \\ 
\;\;\;\;\;\;\;\;\;\;\;\;\;\;\;\;\;\;\;\;\;\;\;\;\;\; \left\{\prod_{t=1}^T f^{\mathbb{P}}(\mathcal{P}_{t}|\mathcal{P}_{t-1},M_{t-1}, k_{\infty}^{\mathbb{Q}}, g^{\mathbb{Q}},\Sigma_{\mathcal{P}},\ell_K,\lambda_{1,2},\widehat{\sigma}_K)\right\} \times \pi(\theta) 
\end{array}
\end{equation}
As the above posterior is not available in closed form, methods such as MCMC can be used to draw samples from it %to calculate expectations of interest, $E[g(\theta)|Y,M,\widehat{\sigma}_K]$, provided that they exist, 
using Monte Carlo. Yet, %note that 
the MCMC output is not %guaranteed
assured to lead to %accurate 
precise Monte Carlo calculations since the corresponding Markov chain may have poor mixing and convergence properties, what leads to highly autocorrelated samples.

It is thus necessary to devise a suitable MCMC algorithm that does not exhibit such unfavourable traits. For further details regarding its construction, see the related appendix in \cite{DTKK2021b}. %characteristics. Various studies (see, for example, \cite{Chib09} and \cite{Bauer18}), note a substantial improvement to the quality of the MCMC output if a Gibbs scheme is adopted, where the parameters are updated in blocks, some of them being full Gibbs steps. For the remaining blocks, independence samplers may be constructed, using the maximum likelihood estimate as the mean in respective proposal density and negative inverse of the corresponding Hessian as its covariance.
Such an MCMC scheme is shown in Algorithm \ref{tab:mcmc}.
\begin{algorithm}[!ht]
\begin{flushleft}
%\hrule
%\medskip
{\itshape \small
\vspace{0.2cm}
Initialize all values of $\theta$. Then at each iteration of the algorithm: \vspace{0.2cm}
\begin{itemize}
\item[(a)] Update $\sigma_e^2$ from its full conditional distribution that can be shown to be an Inverse Gamma distribution with parameters $\tilde{\alpha}/2$ and $\tilde{\beta}/2$, such that $\tilde{\alpha}$ is $\alpha+T(J-R)$ and $\tilde{\beta}$ is $\beta+\sum^{T}_{t=0}\|\hat{e}_t\|^2$, where $\alpha=\beta=0$, since prior is assumed diffuse, $\hat{e}_t$ is a time-$t$ residual from \eqref{Qmodel}, and $\|\cdot\|^2$ is Euclidean norm squared.
\item[(b)] Update $\Sigma_{\mathcal{P}}$ using an independence sampler based on the MLE and the Hessian obtained before running the MCMC, using multivariate t-distribution with 5 degrees of freedom as proposal distribution.
\item[(c)] Update $(k_{\infty}^{\mathbb{Q}},g^{\mathbb{Q}})$ in a similar manner to (b).
\item[(d)] Update $(\ell_K,\lambda_{1,2})$ in a similar manner to (b).
\end{itemize}}
%\medskip
%\hrule
\end{flushleft}
\caption{MCMC scheme for Gaussian Affine Term Structure Models with unspanned nonlinear macros}
\label{tab:mcmc}
\end{algorithm}

\section{Adaptive Tempering}
\label{appendix:adaptivetempering}

The aim of adaptive tempering is to smooth peaked likelihoods. For additional information, see the corresponding appendix to \cite{DTKK2021b}. %It is achieved by bridging two successive targets via an intermediate target sequence. The idea is to modify the sequence of target distributions so that is evolves from the prior to the posterior more smoothly. See \cite{jasra2011} and \cite{Kantas2014} for details.
Implementation of the IBIS scheme with hybrid adaptive tempering steps is outlined in Algorithm \ref{tab:ibisat}. It is important to note that, unlike it is presented in Algorithm \ref{tab:ibis} for the general IBIS case, in the specific case we are dealing here with we initialize the particles by drawing from the posterior  $\pi(\theta|Y_{0:t-1},M_{0:t-2},\widehat{\sigma}_K)$ instead of the prior $\pi(\theta)$. This is done in-sample based on training data, as detailed in Section \ref{subsec:yieldsmacros}.

Although it is straightforward to implement step 4(b)$iv$ in Algorithm \ref{tab:ibisat} for an independence sampler, adjustments are necessary for a full Gibbs step. It is especially the case for $\sigma_e^2$ in step (a) in Algorithm \ref{tab:mcmc}, see Appendix \ref{appendix:MCMC}. Implementation details are the same as in the corresponding appendix to the paper we mention above. %We begin by observing that, assuming tempering parameter $\phi$, the following holds
% $$
% \left[f(\hat{e}_T|\sigma_e^2)\right]^{\phi}=\frac{\exp{\left(-\frac{\phi}{2\sigma_e^2} \|\hat{e}_T\|^2 \right)}}{\left[(2\pi)^{J-N}\sigma_e^{2(J-N)}(J-N)\right]^{\frac{\phi}{2}}}
% $$
% for $\hat{e}_T$ which is a time-$T$ residual from \eqref{Qmodel}, $f(\hat{e}_T|\sigma_e^2)$ is the associated likelihood and $\|\cdot\|^2$ denotes Euclidean norm squared. Then for $\hat{e}_{0:T}$ we let
% $$
% f(\hat{e}_{0:T}|\sigma_e^2,\phi)=f(\hat{e}_{0:T-1}|\sigma_e^2)\times\left[f(\hat{e}_T|\sigma_e^2)\right]^{\phi}
% $$
% be the corresponding tempered likelihood. Combined with an $IG\left(\alpha/2,\;\beta/2\right)$ prior that is proportional to
% $$
% p(\sigma_e^2|\alpha,\beta) \propto \left(\sigma_e^2\right)^{-\frac{\alpha}{2}-1}\times \exp{\left(-\frac{\beta}{2\sigma_e^2}\right)} 
% $$
% we get tempered posterior as
% $$
% \pi(\sigma_e^2|\hat{e}_{0:T},\alpha,\beta,\phi) \propto f(\hat{e}_{0:T}|\sigma_e^2,\phi)\times p(\sigma_e^2|\alpha,\beta)
% $$
% and eventually we arrive at
% $$
% \pi(\sigma_e^2|\hat{e}_{0:T},\alpha,\beta,\phi) \propto \left(\sigma_e^2\right)^{-\frac{1}{2}\left[\alpha+(T-1+\phi)(J-N)\right]-1}\times \exp{\left(-\frac{\beta+\phi\|\hat{e}_T\|^2 + \sum_{t=0}^{T-1}\|\hat{e}_t\|^2}{2\sigma_e^2}\right)} 
% $$
% what assuming 
% \begin{eqnarray}
% \tilde{\alpha}&=&\alpha+(T-1+\phi)(J-N) \\
% \tilde{\beta}&=&\beta+\phi\|\hat{e}_T\|^2 + \sum_{t=0}^{T-1}\|\hat{e}_t\|^2
% \end{eqnarray}
% leads us to an $IG(\tilde{\alpha}/2,\;\tilde{\beta}/2)$ posterior for $\sigma_e^2$.
%
\begin{algorithm}[!ht]
\begin{flushleft}
%\hrule
%\medskip
{\itshape \small
\vspace{0.2cm}
Initialize $N_{\theta}$ particles by drawing independently $\theta_{i}\sim \pi(\theta|Y_{0:t-1},M_{0:t-2},\widehat{\sigma}_K)$ with importance weights $\omega_{i}=1$, $i=1,\dots,N_{\theta}$. For $t,\dots,T$ and each time for all $i$:\vspace{0.2cm}
\begin{itemize}
\item[1] Set $\omega_{i}'=\omega_{i}$. 
\item[2] Calculate the incremental weights from
$$
u_t(\theta_{i},Y_{0:t-1},M_{0:t-1})=f\big(Y_{t}|Y_{0:t-1},M_{0:t-1},\theta_{i},\widehat{\sigma}_K)
$$
\item[3] Update the importance weights $\omega_{i}$ to $\omega_{i}u_t(\theta_{i},Y_{0:t-1},M_{0:t-1})$.
\item[4] If degeneracy criterion ESS($\omega$) is triggered, perform the following sub-steps:
\begin{itemize}
\item[(a)] Set $\phi=0$ and $\phi'=0$.
\item[(b)] While $\phi<1$
\begin{itemize}
\item[i.] If degeneracy criterion ESS($\omega''$) is not triggered, where $\omega_{i}''=\omega_{i}'[u_t(\theta_{i},Y_{0:t-1},M_{0:t-1})]^{1-\phi'}$, set $\phi=1$, otherwise find $\phi\in[\phi',1]$ such that ESS($\omega'''$) is greater than or equal to the trigger, where $\omega_{i}'''=\omega_{i}'[u_t(\theta_{i},Y_{0:t-1},M_{0:t-1})]^{\phi-\phi'}$, for example using bisection method, see \cite{Kantas2014}.
\item[ii.] Update the importance weights $\omega_{i}$ to $\omega_{i}'[u_t(\theta_{i},Y_{0:t-1},M_{0:t-1})]^{\phi-\phi'}$.
\item[iii.] Resample: Sample with replacement $N_{\theta}$ times from the set of $\theta_{i}$s according to their weights $\omega_i$. The weights are then reset to one.
\item[iv.] Jitter: Replace $\theta_i$s with $\tilde{\theta}_i$s by running MCMC chains with each $\theta_i$ as input and $\tilde{\theta}_i$ as output, using likelihood given by $f(Y_{0:t-1}|M_{0:t-2},\theta_{i},\widehat{\sigma}_K)[f\big(Y_{t}|M_{t-1},\theta_{i},\widehat{\sigma}_K)]^{\phi}$. Set $\theta_i=\tilde{\theta}_i$. 
\item[v.] Calculate the incremental weights from
$$
u_t(\theta_{i})=f\big(Y_{t}|Y_{0:t-1},M_{0:t-1},\theta_{i},\widehat{\sigma}_K)
$$
\item[vi.] Set $\omega_{i}'=\omega_{i}$ and $\phi'=\phi$.
\end{itemize}
\end{itemize} 
\end{itemize}}
%\medskip
%\hrule
\end{flushleft}
\caption{IBIS algorithm with hybrid adaptive tempering for Gaussian Affine Term Structure Models with unspanned nonlinear macros}
\label{tab:ibisat}
\end{algorithm}

\section{Tuning the Gaussian Process}
\label{appendix:GPcalibration}

Since we view $\sigma_K$ as tuning parameter %For identification purposes, we calibrate $\sigma_K$ 
we tune it in-sample and fix out-of-sample at $\widehat{\sigma}_K$. Details about the underlying data are in Section \ref{subsec:yieldsmacros}. To that end, in a manner similar to this in the corresponding appendix in \cite{DTKK2021a}, we follow a multi-step process which is entirely based on in-sample data. First, as in the base case in Section \ref{subsec:basecase}, we estimate by maximum likelihood a yields-only DTSM where $N=3$ and, out of $\lambda_{\mathcal{P}}$, only $\lambda_{1,2}$ is left unrestricted to match the risk price restrictions we adopt in this paper. Resulting MLEs let us then obtain $\hat{s}_t$, $t=1,...,\widetilde{T}$, where $\widetilde{T}$ refers to in-sample period, from \eqref{P2s}.

Second, we formulate an amended version of the likelihood in \eqref{likelihoodlambda12}, using only its $\mathbb{P}$-likelihood components $f^{\mathbb{P}}(\cdot)$ modified in the following way 
\begin{equation}\label{likelihoodcalibrate}
\tilde{f}(\widetilde{Y}|\widetilde{M},\theta,\hat{k}_{\infty}^{\mathbb{Q}}, \hat{g}^{\mathbb{Q}}, \widehat{\Sigma}_{\mathcal{P}},\hat{\lambda}_{1,2})= \left\{\prod_{t=1}^{\widetilde{T}} \tilde{f}^{\mathbb{P}}(\mathcal{P}_{t}|\mathcal{P}_{t-1},M_{t-1}, \hat{k}_{\infty}^{\mathbb{Q}}, \hat{g}^{\mathbb{Q}},\widehat{\Sigma}_{\mathcal{P}},\ell_K,\hat{\lambda}_{1,2},c)\right\}
\end{equation}
where $\widetilde{Y}$ and $\widetilde{M}$ refer to in-sample data, $\hat{k}_{\infty}^{\mathbb{Q}}$, $\hat{g}^{\mathbb{Q}}$, $\widehat{\Sigma}_{\mathcal{P}}$ and $\hat{\lambda}_{1,2}$ are the MLEs from the first step, and $\theta=(\ell_K,c)$, with scalar $c>0$, consists of parameters we estimate by maximum likelihood next. However, before that we parametrize $\sigma_K$ in \eqref{likelihoodcalibrate} as
\begin{equation}\label{sigmaKhat}
\sigma_K= c \; \sqrt{diagv\left[Var(\hat{s})\right]}
\end{equation}
where $diagv\left[Var(\hat{s})\right]$ is a $3\times 1$ (in accordance with exposition in this paper where we choose $G=3$) vector including diagonal elements of the covariance matrix $Var(\hat{s})$ for $\hat{s} = \left[\hat{s}_1,\dots,\hat{s}_{\widetilde{T}} \right]$, and $\hat{s}_t$, $t=1,\dots,\widetilde{T}$, are practically as in the first step.

Third, as in Section \ref{subsec:likelihoodGaussian}, we proceed with the the log-likelihood representation of \eqref{likelihoodcalibrate}, similar to \eqref{loglikP}, which we maximize. Consequently, it lets us fix $\sigma_K$ out-of-sample at $\widehat{\sigma}_K$, which we calculate from \eqref{sigmaKhat} with $\hat{c}$ being the MLE of $c$ from this last step and $Var(\hat{s})$ remains unchanged. %Calibration results for different models are in Table \ref{table:SIGs}.

\section{Linear Model with Macros}
\label{appendix:LinearModel}

It is straightforward to extend the estimation framework of \cite{Bauer18}, and consequently this in \cite{DTKK2021b}, to incorporate in a linear manner unspanned macros which are assumed exogenous. % according to \eqref{PmodelExog}.
It is only the way we handle the $\mathbb{P}$-dynamics of $\mathcal{P}_t$ in \eqref{PmodelExog} what needs to be adjusted to
\begin{equation}
\label{PmodelExogLin}
\mathcal{P}_t = \mu_{\mathcal{P}}^{\mathbb{P}} + \Phi_{\mathcal{P}}^{\mathbb{P}}\mathcal{P}_{t-1} + \Phi_{\mathcal{P}M}^{\mathbb{P}} M_{t-1} +\Sigma_{\mathcal{P}} \varepsilon_{t}^{\mathcal{P}}
\end{equation}
where $\Phi_{\mathcal{P}M}^{\mathbb{P}}$ is $(N \times R)$ matrix, which represents the feedback from $M_{t-1}$ to $\mathcal{P}_t$. 

Following sections C.1 and C.2 in Online Appendix to \cite{Bauer18} and details from \cite{Lutkepohl2005}, to consider $\Phi_{\mathcal{P}M}^{\mathbb{P}}$ coefficients next to $M_{t-1}$ in \eqref{PmodelExogLin} in estimation, it suffices to tackle them jointly with $\lambda_{\mathcal{P}}$, or in our case with $\lambda_{1,2}$ only, to match the risk price restrictions chosen for the nonlinear case in Section \ref{subsec:likelihoodGaussian}. Adopting notation from Online Appendix to \cite{Bauer18} to ours where necessary, we can rewrite \eqref{PmodelExogLin} in vector form as
$$
X = B Z+U
$$
where
$X=\left[\mathcal{P}_1,\dots,\mathcal{P}_T \right]$, $U = \left[u_1,\dots,u_{T}\right]$, with adapted $u_t =\Sigma_{\mathcal{P}}\varepsilon_t^{\mathcal{P}}$, $t=1,\dots,T$, $Z = \left[Z_{0},\dots,Z_{(T-1)}\right]$, with modified $Z_{t} = \left[1,\mathcal{P}_t',M_t'\right]'$, $t=0,\dots,T-1$, and amended $B=\left(\mu_{\mathcal{P}}^{\mathbb{P}},\Phi_{\mathcal{P}}^{\mathbb{P}},\Phi_{\mathcal{P}M}^{\mathbb{P}}\right)$.

Then, linear constraints after \cite{Bauer18} are following
$$
\beta = vec(B) = \lambda + r = S\lambda_{\gamma} + r
$$
where in our case $S$ is a $[N(N+R+1)]\times [NR+1]$ selection matrix of zeros and ones, $\lambda_{\gamma}$ is a $[NR+1]\times 1$ vector with $\lambda_{1,2}$ and those elements of $\Phi_{\mathcal{P}M}^{\mathbb{P}}$ we decide to leave unrestricted. For example, if our goal it to compare a linear model with a corresponding nonlinear case where $G=2$ and there is no $GP$ included in the first equation of \eqref{PmodelExogGP}, to allow for a meaningful comparison of results we would restrict the first row in $\Phi_{\mathcal{P}M}^{\mathbb{P}}$ to zeros, in a similar fashion to restricting risk prices in $\lambda_{\mathcal{P}}$. Finally, $r=vec\left[\mu_{\mathcal{P}}^{\mathbb{Q}},\Phi_{\mathcal{P}}^{\mathbb{Q}},0_{NR\times 1}\right]$. For clarity, $\lambda$ contains all elements of $\lambda_{\gamma}$, as well as $N(N+R+1)-[NR+1]$ zeros, and in our case $N=3$ and $R=1$.

After the above modifications the rest is straightforward to conclude and one can easily follow in the footsteps of \cite{Bauer18} for a Bayesian framework, as well as \cite{DTKK2021b} for a corresponding sequential implementation thereof, to eventually arrive at an almost complete estimation framework for a linear model with macros which is comparable to the setup we develop in this paper for the nonlinear case. What is still missing though is the prior specification for $\Phi_{\mathcal{P}M}^{\mathbb{P}}$. This we choose to be non-informative and thus assign independent normal distribution with zero mean and large variance to each element thereof. 

\end{appendices}
\newpage

% GRO

% CPI

% GRO

\newpage

\begin{table}[!ht]\scriptsize \caption[Explanatory power gains from macroeconomic variables when fitting excess bond returns]{Explanatory power gains from macroeconomic variables, $CPI$ and $GRO$, when fitting excess bond returns, measured via $\bar{R}^2$ at 1-month prediction horizon - period: January 1985 - end of 2017.}
\label{table:deltaR2adj0}
\begin{center}
\begin{tabular}{*{7}{p{1.5cm}}}%{ccccccc}
\toprule
\multicolumn{1}{p{1.5cm}}{$\bf n$} &\multicolumn{1}{p{1.5cm}}{\bf 2Y} &\multicolumn{1}{p{1.5cm}}{\bf 3Y} & \multicolumn{1}{p{1.5cm}}{\bf 4Y} & \multicolumn{1}{p{1.5cm}}{\bf 5Y} & \multicolumn{1}{p{1.5cm}}{\bf 7Y} & \multicolumn{ 1}{p{1.5cm}}{\bf 10Y} \\
\midrule 
\multicolumn{1}{c}{} &\multicolumn{ 6}{l}{\textbf{$\bar{R}^2(\%):rx_{t,t+1}^n=a+b'\mathcal{P}_t+e_t$}} \\
\midrule
 &       4.26 &       3.29 &       3.08 &       2.74 &       2.77 &       3.55 \\
\midrule 
\multicolumn{1}{c}{} & \multicolumn{ 6}{l}{\textbf{$\Delta_{(0)}\bar{R}^2(\%):rx_{t,t+1}^n=a+b'\mathcal{P}_t+cM_{t-1}+e_t$}} \\
\midrule
$CPI$ &   -0.24 &      -0.24 &      -0.22 &      -0.16 &       0.05 & $0.10^{*}$ \\
$GRO$ & $1.95^{***}$ & $1.47^{***}$ & $0.56^{*}$ &  0.08 &   -0.21 &   -0.17 \\
% $F8$ &   -0.25 &      -0.24 &      -0.19 &      -0.24 &      -0.24 &      -0.23 \\
\midrule 
\multicolumn{1}{c}{} & \multicolumn{ 6}{l}{\textbf{$\bar{R}^2(\%):rx_{t,t+1}^n=a+b'\mathcal{P}_t+cM_{t-1}+e_t$}} \\
\midrule
$CPI$ &   4.03 &       3.06 &       2.86 &       2.58 &       2.81 &       3.65 \\
$GRO$ &  6.13 &       4.71 &       3.61 &       2.82 &       2.56 &       3.38 \\
% $F8$ &  4.02 &       3.06 &       2.89 &       2.50 &       2.53 &       3.32 \\
\bottomrule
\end{tabular}
\end{center}
\noindent{\caption*{\tiny This table reports in-sample $\bar{R}^2$ in $\%$ across alternative regression specifications at $h$ = 1-month prediction horizon. The explained variables are different (by maturities) excess bond returns. The explanatory variables are the principal components $\mathcal{P}_t$ and observed (lagged) macros $M_{t-1}$, in particular $CPI$ and $GRO$. In-sample $\bar{R}^2$ values are obtained in a similar manner to the out-of-sample $R^2$ measure of \cite{Campbell08} but using in-sample fit instead of out-of-sample forecasts, and incorporating penalty adjustment. In particular, $\bar{R}^2$ in the top panel (all highly statistically significant hence not denoted) measures explanatory power gains from using principal components on top of the in-sample average to fit excess bond returns, whereas $\bar{R}^2$ in the bottom panel (all highly statistically significant hence not denoted) measures explanatory power gains from using principal components and $M_{t-1}$ on top of the in-sample average to fit excess bond returns. Further, $\Delta_{(0)}$ next to $\bar{R}^2$ in the mid panel means that the latter measures explanatory power gains from using macro $M_{t-1}$ on top of the in-sample average and the principal components to do the same. Positive values of this statistic imply that there is explanatory power gain from adding extra variables. Statistical significance is measured using a one-sided Diebold-Mariano statistic with Clark-West adjustment, based on Newey-West standard errors. * denotes significance at 10\%, ** significance at 5\% and *** significance at 1\% level. The sample period is January 1985 to end of 2017.}}
\end{table}

\begin{table}[!ht]\scriptsize \caption[Explanatory power gains from nonlinear macros estimated using model $GP_{111}$, when fitting excess bond returns]{Explanatory power gains from nonlinear macros estimated using model $GP_{111}$, when  fitting excess bond returns, measured via $\bar{R}^2$ at 1-month prediction horizon - period: January 1985 - end of 2017.}
\label{table:deltaR2adj111}
\begin{center}
\begin{tabular}{*{7}{p{1.5cm}}}%{ccccccc}
\toprule
\multicolumn{1}{p{1.5cm}}{$\bf n$} &\multicolumn{1}{p{1.5cm}}{\bf 2Y} &\multicolumn{1}{p{1.5cm}}{\bf 3Y} & \multicolumn{1}{p{1.5cm}}{\bf 4Y} & \multicolumn{1}{p{1.5cm}}{\bf 5Y} & \multicolumn{1}{p{1.5cm}}{\bf 7Y} & \multicolumn{ 1}{p{1.5cm}}{\bf 10Y} \\
\midrule 
\multicolumn{1}{c}{} &\multicolumn{ 6}{l}{\textbf{$\bar{R}^2(\%):rx_{t,t+1}^n=a+b'\mathcal{P}_t+e_t$}} \\
\midrule
 &       4.26 &       3.29 &       3.08 &       2.74 &       2.77 &       3.55 \\
\midrule 
\multicolumn{1}{l}{$GP_{111}$} &\multicolumn{6}{l}{\textbf{$\Delta_{(3)}\bar{R}^2(\%):rx_{t,t+1}^n=a+b'\mathcal{P}_t+c\;v^{(3)}(M_{t-1})+e_t$}} \\
\midrule
$CPI$ & -0.14 &       0.17 & $0.44^{*}$ & $0.71^{**}$ & $1.48^{**}$ & $2.29^{***}$ \\
$GRO$ & -0.24 &      -0.11 &       0.04 &       0.26 & $1.19^{*}$ & $2.19^{**}$ \\
% $F8$ & -0.07 &       0.06 & $0.29^{*}$ & $0.41^{*}$ & $0.85^{**}$ & $1.02^{**}$ \\
\midrule    
\multicolumn{1}{l}{$GP_{111}$} &\multicolumn{6}{l}{\textbf{$\Delta_{(2)3}\bar{R}^2(\%):rx_{t,t+1}^n=a+b'\mathcal{P}_t+c'[v^{(2)}(M_{t-1}),v^{(3)}(M_{t-1})]'+e_t$}} \\
\midrule
$CPI$ & $1.01^{**}$ & $0.42^{*}$ &       0.17 &      -0.12 &      -0.20 &      -0.25 \\
$GRO$ & $1.20^{**}$ & $0.52^{*}$ &       0.11 &      -0.15 &      -0.25 &      -0.25 \\
% $F8$ & $0.30^{*}$ &   0.14 &      -0.02 &      -0.10 &      -0.12 &      -0.25 \\
\midrule    
\multicolumn{1}{l}{$GP_{111}$} &\multicolumn{6}{l}{\textbf{$\Delta_{(1)23}\bar{R}^2(\%):rx_{t,t+1}^n=a+b'\mathcal{P}_t+c'[v^{(1)}(M_{t-1}),v^{(2)}(M_{t-1}),v^{(3)}(M_{t-1})]'+e_t$}} \\
\midrule
$CPI$ & $0.66^{**}$ & $0.58^{**}$ & $0.59^{**}$ & $0.64^{**}$ &       0.07 &      -0.13 \\
$GRO$ & $2.91^{***}$ & $2.86^{***}$ & $2.74^{***}$ & $2.17^{***}$ & $1.27^{***}$ & $0.34^{*}$ \\
% $F8$ & $1.38^{***}$ & $1.12^{**}$ & $0.61^{**}$ & $0.47^{**}$ &      -0.12 &      -0.25 \\
\midrule    
\multicolumn{1}{l}{$GP_{111}$} &\multicolumn{6}{l}{\textbf{$\Delta_{(123)}\bar{R}^2(\%):rx_{t,t+1}^n=a+b'\mathcal{P}_t+c'[v^{(1)}(M_{t-1}),v^{(2)}(M_{t-1}),v^{(3)}(M_{t-1})]'+e_t$}} \\
\midrule
$CPI$ & $1.51^{***}$ & $1.16^{***}$ & $1.20^{**}$ & $1.23^{***}$ & $1.34^{**}$ & $1.91^{***}$ \\
$GRO$ & $3.85^{***}$ & $3.25^{***}$ & $2.88^{***}$ & $2.28^{***}$ & $2.20^{***}$ & $2.27^{***}$ \\
% $F8$ & $1.60^{***}$ & $1.31^{***}$ & $0.87^{***}$ & $0.77^{***}$ & $0.61^{**}$ & $0.53^{**}$ \\
\midrule    
\multicolumn{1}{l}{$GP_{111}$} &\multicolumn{6}{l}{\textbf{$\bar{R}^2(\%):rx_{t,t+1}^n=a+b'\mathcal{P}_t+c'[v^{(1)}(M_{t-1}),v^{(2)}(M_{t-1}),v^{(3)}(M_{t-1})]'+e_t$}} \\
\midrule
$CPI$ & 5.71 &       4.40 &       4.23 &       3.93 &       4.07 &       5.39 \\
$GRO$ & 7.94 &       6.43 &       5.86 &       4.95 &       4.90 &       5.73 \\
% $F8$ & 5.78 &       4.55 &       3.92 &       3.49 &       3.36 &       4.05 \\
\bottomrule
\end{tabular}
\end{center}
\noindent{\caption*{\tiny This table reports in-sample $\bar{R}^2$ in $\%$ across alternative regression specifications at $h$ = 1-month prediction horizon. The explained variables are different (by maturities) excess bond returns. The explanatory variables are the principal components $\mathcal{P}_t$ and the estimated (via posterior mean) nonlinear macros $v(M_{t-1})$ from model $GP_{111}$, for $CPI$ and $GRO$. In-sample $\bar{R}^2$ values are obtained in a similar manner to the out-of-sample $R^2$ measure of \cite{Campbell08} but using in-sample fit instead of out-of-sample forecasts, and incorporating penalty adjustment. In particular, $\bar{R}^2$ in the top panel (all highly statistically significant hence not denoted) measures explanatory power gains from using principal components on top of the in-sample average to fit excess bond returns, whereas $\bar{R}^2$ in the bottom panel (all highly statistically significant hence not denoted) measures explanatory power gains from using principal components and $v^{(1)}(M_{t-1})$, $v^{(2)}(M_{t-1})$ and $v^{(3)}(M_{t-1})$ on top of the in-sample average to fit excess bond returns. Further, $\Delta_{(3)}$ next to $\bar{R}^2$ in the second panel means that the latter measures explanatory power gains from using nonlinear macro $v^{(3)}(M_{t-1})$ estimated using model $GP_{111}(M)$ on top of the in-sample average and the principal components to do the same. Next, $\Delta_{(2)3}$ in the third panel means that the associated $\bar{R}^2$ measures explanatory power gains from using to that end the nonlinear macro $v^{(2)}(M_{t-1})$ on top of the in-sample average, the principal components and nonlinear macro $v^{(3)}(M_{t-1})$. Then, $\Delta_{(1)23}$ in the fourth panel means that the associated $\bar{R}^2$ measures explanatory power gains from using to that end the nonlinear macro $v^{(1)}(M_{t-1})$ on top of the in-sample average, the principal components and nonlinear macros $v^{(2)}(M_{t-1})$ and $v^{(3)}(M_{t-1})$. Finally, $\Delta_{(123)}$ next to $\bar{R}^2$ in the fifth panel means that is measures the joint effect of including all latent factors on top of the remaining explanatory variables to fit excess bond returns. Positive values of this statistic imply that there is explanatory power gain from adding extra variables. Statistical significance is measured using a one-sided Diebold-Mariano statistic with Clark-West adjustment, based on Newey-West standard errors. * denotes significance at 10\%, ** significance at 5\% and *** significance at 1\% level. The sample period is January 1985 to end of 2017.}}
\end{table}

\begin{table}[!ht]\scriptsize \caption[Out-of-sample statistical performance of bond excess return forecasts against the EH]{Out-of-sample statistical performance of bond excess return forecasts against the EH, measured via $R_{os}^2$ at 1-month prediction horizon - period: January 1985 - end of 2018.}
\label{table:R2OS}
\begin{center}
\begin{tabular}{lllllll}%{ccccccc}
\toprule % \bf h=1m \backslash
\multicolumn{1}{l}{$\bf n$} &\multicolumn{1}{m{1.3cm}}{\bf 2Y} &\multicolumn{1}{m{1.3cm}}{\bf 3Y} & \multicolumn{1}{m{1.3cm}}{\bf 4Y} & \multicolumn{1}{m{1.3cm}}{\bf 5Y} & \multicolumn{1}{m{1.3cm}}{\bf 7Y} & \multicolumn{ 1}{m{1.3cm}}{\bf 10Y} \\
\midrule                                              
$M_0$\hspace{0.75cm} &     -0.04  &     -0.06  &     -0.05  &     -0.04  &     -0.02  &    -0.01** \\

$M_1$\hspace{0.75cm} &      0.01  &     0.03** &      0.03* &      0.02  &      0.02* &     0.04** \\
\midrule  
\multicolumn{ 7}{c}{\textbf{$CPI$}} \\
\midrule
$LM_{111}$\hspace{0.5cm} &     0.03** &    0.04*** &     0.03** &     0.03** &     0.03** &    0.03*** \\
$GP_{111}$\hspace{0.5cm} &      0.02* &     0.04** &     0.03** &      0.03* &      0.03* &    0.05*** \\
\midrule
$LM_{011}$\hspace{0.5cm} &     0.03** &    0.05*** &     0.04** &     0.03** &     0.03** &    0.04*** \\
$GP_{011}$\hspace{0.5cm} &     0.03** &    0.04*** &     0.03** &     0.03** &      0.03* &    0.04*** \\
\midrule
$LM_{110}$\hspace{0.5cm} &    0.03** & 0.03*** & 0.03** & 0.02** & 0.02** & 0.03*** \\
$GP_{110}$\hspace{0.5cm} &    0.06*** & 0.06*** & 0.05*** & 0.04** & 0.04** & 0.04*** \\
% \midrule
% $LM_{101}$\hspace{0.5cm} &    0.04** & 0.04*** & 0.03*** & 0.03** & 0.03** & 0.04*** \\
% $GP_{101}$\hspace{0.5cm} &    0.03** & 0.05*** & 0.04** & 0.03** & 0.03** & 0.05*** \\
% \midrule
% $LM_{100}$\hspace{0.5cm} &    0.03** & 0.03** & 0.03** & 0.02* & 0.03** & 0.03*** \\
% $GP_{100}$\hspace{0.5cm} &    0.03* & 0.04** & 0.04** & 0.03* & 0.03* & 0.04*** \\
\midrule
$LM_{010}$\hspace{0.5cm} &    0.03* & 0.04*** & 0.04** & 0.03** & 0.03** & 0.04*** \\
$GP_{010}$\hspace{0.5cm} &    0.05*** & 0.05*** & 0.04*** & 0.04** & 0.03** & 0.04*** \\
\midrule
$LM_{001}$\hspace{0.5cm} &      0.03* &     0.04** &     0.03** &      0.03* &      0.03* &    0.04*** \\
$GP_{001}$\hspace{0.5cm} &     0.03** &    0.04*** &     0.04** &     0.03** &      0.03* &    0.04*** \\
\midrule
\multicolumn{ 7}{c}{\textbf{$GRO$}} \\
\midrule
$LM_{111}$\hspace{0.5cm} &      -0.19 &      -0.03 &      -0.03 &      -0.04 &      -0.01 &       0.01 \\
$GP_{111}$\hspace{0.5cm} &     -0.03  &      0.01* &      0.01* &      0.01  &      0.01  &      0.01* \\
\midrule
$LM_{011}$\hspace{0.5cm} &      0.02  &     0.05** &    0.04*** &     0.03** &      0.02* &     -0.02  \\
$GP_{011}$\hspace{0.5cm} &      0.03* &     0.05** &     0.03** &     0.02** &      0.01  &      0.00  \\
\midrule
$LM_{110}$\hspace{0.5cm} &    -0.22 & -0.05 & -0.05 & -0.07 & -0.04 & 0.00 \\
$GP_{110}$\hspace{0.5cm} &    -0.04 & 0.00 & 0.00 & 0.00 & 0.00 & 0.00 \\
% \midrule
% $LM_{101}$\hspace{0.5cm} &    -0.19 & -0.05 & -0.07 & -0.10 & -0.05 & 0.00 \\
% $GP_{101}$\hspace{0.5cm} &    -0.03 & 0.00 & 0.00 & -0.01 & 0.00 & 0.02* \\
% \midrule
% $LM_{100}$\hspace{0.5cm} &    -0.22 & -0.08 & -0.09 & -0.12 & -0.07 & -0.01 \\
% $GP_{100}$\hspace{0.5cm} &    -0.07 & -0.02 & -0.02 & -0.02 & -0.01 & 0.01* \\
\midrule
$LM_{010}$\hspace{0.5cm} &   0.01 & 0.05** & 0.03** & 0.02** & 0.02* & -0.01 \\
$GP_{010}$\hspace{0.5cm} &   0.03* & 0.05** & 0.03** & 0.02** & 0.01* & 0.00 \\
\midrule
$LM_{001}$\hspace{0.5cm} &      0.03* &     0.04** &     0.04** &     0.04** &      0.03* &     0.03** \\
$GP_{001}$\hspace{0.5cm} &     0.03** &    0.04*** &     0.04** &     0.03** &     0.03** &    0.04*** \\
\bottomrule
\end{tabular}
\end{center}
\noindent{\caption*{\tiny This table reports out-of-sample $R^2$ across alternative models at $h$ = 1-month  prediction horizon. The forecasting models used are DTSM with either alternative risk price restrictions or different number of $GP$ outputs with various macros or just other macros. $R^2$ values are generated using the out-of-sample $R^2$ measure of \cite{Campbell08}. In particular, out-of-sample $R^2$ measures the predictive accuracy of bond excess return forecasts relative to the EH benchmark. The EH implies the historical mean being the optimal forecast of excess returns. Positive values of this statistic imply that the forecast outperforms the historical mean forecast and suggests evidence of time-varying return predictability. Statistical significance is measured using a one-sided Diebold-Mariano statistic with Clark-West adjustment, based on Newey-West standard errors. * denotes significance at 10\%, ** significance at 5\% and *** significance at 1\% level. The in-sample period is January 1985 to end of 2007, and the out-of-sample period starts in January 2008 and ends in end of 2018.}}
\end{table}

\begin{table}[!ht]\scriptsize \caption[Out-of-sample statistical performance of bond excess return forecasts against the corresponding linear model]{Out-of-sample statistical performance of bond excess return forecasts against the corresponding linear model, measured via $R_{os}^2$ at 1-month prediction horizon - period: January 1985 - end of 2018.}
\label{table:R2OSLM}
\begin{center}
\begin{tabular}{lllllll}%{ccccccc}
\toprule % \bf h=1m \backslash
\multicolumn{1}{l}{$\bf n$} &\multicolumn{1}{m{1.3cm}}{\bf 2Y} &\multicolumn{1}{m{1.3cm}}{\bf 3Y} & \multicolumn{1}{m{1.3cm}}{\bf 4Y} & \multicolumn{1}{m{1.3cm}}{\bf 5Y} & \multicolumn{1}{m{1.3cm}}{\bf 7Y} & \multicolumn{ 1}{m{1.3cm}}{\bf 10Y} \\
\midrule  
\multicolumn{ 7}{c}{\textbf{$CPI$}} \\
\midrule
$GP_{111}$\hspace{0.5cm}  &     -0.01  &      0.00  &      0.00  &      0.00  &      0.01  &      0.02* \\
$GP_{011}$\hspace{0.5cm}  &       0.00 &       0.00 &       0.00 &       0.00 &      -0.01 &       0.00 \\
$GP_{110}$\hspace{0.5cm}  &     0.03** & 0.03** & 0.02* & 0.01 & 0.01 & 0.02* \\
% $GP_{101}$\hspace{0.5cm}  &    -0.01 & 0.01 & 0.01 & 0.01 & 0.01 & 0.01* \\
% $GP_{100}$\hspace{0.5cm}  &   0.00 & 0.01 & 0.01 & 0.01 & 0.01 & 0.01* \\
$GP_{010}$\hspace{0.5cm}  &   0.02** & 0.01 & 0.01 & 0.00 & 0.00 & 0.01 \\
$GP_{001}$\hspace{0.5cm}  &      0.01  &      0.01* &      0.01  &      0.01  &      0.00  &     -0.01  \\
\midrule
\multicolumn{ 7}{c}{\textbf{$GRO$}} \\
\midrule
$GP_{111}$\hspace{0.5cm}  &      0.13* &      0.04  &      0.04  &      0.05* &      0.02  &      0.01  \\
$GP_{011}$\hspace{0.5cm}  &       0.01 &      -0.01 &      -0.01 &      -0.01 &      -0.01 &       0.02 \\
$GP_{110}$\hspace{0.5cm}  &      0.15** & 0.05* & 0.05 & 0.06* & 0.03 & 0.00 \\
% $GP_{101}$\hspace{0.5cm}  &     0.13* & 0.05* & 0.06 & 0.08* & 0.05 & 0.02 \\
% $GP_{100}$\hspace{0.5cm}  &     0.12* & 0.05* & 0.07* & 0.09* & 0.06 & 0.02 \\
$GP_{010}$\hspace{0.5cm}  &     0.02 & 0.00 & 0.00 & 0.00 & -0.01 & 0.01 \\
$GP_{001}$\hspace{0.5cm}  &       0.00 &       0.00 &       0.00 &      -0.01 &       0.00 &       0.01 \\
% \midrule
% \multicolumn{ 7}{c}{\textbf{$F8$}} \\
% \midrule
% $GP_{111}$\hspace{0.5cm}  &    0.02*** &      0.01* &      0.01* &      0.02* &      0.01  &      0.01  \\
% $GP_{011}$\hspace{0.5cm}  &       0.01 &       0.01 &       0.00 &       0.00 &       0.00 &      -0.01 \\
% $GP_{001}$\hspace{0.5cm}  &       0.01 &       0.00 &       0.00 &       0.00 &      -0.01 &       0.00 \\
\bottomrule
\end{tabular}
\end{center}
\noindent{\caption*{\tiny This table reports out-of-sample $R^2$ across alternative models at $h$ = 1-month  prediction horizon. The forecasting models used are DTSM with different number of $GP$ outputs and various macros or just other macros. $R^2$ values are generated using the out-of-sample $R^2$ measure of \cite{Campbell08}. In particular, out-of-sample $R^2$ measures the predictive accuracy of bond excess return forecasts relative to the corresponding linear model. Positive values of this statistic imply that the forecast outperforms forecast from the benchmark model. Statistical significance is measured using a one-sided Diebold-Mariano statistic with Clark-West adjustment, based on Newey-West standard errors. * denotes significance at 10\%, ** significance at 5\% and *** significance at 1\% level. The in-sample period is January 1985 to end of 2007, and the out-of-sample period starts in January 2008 and ends in end of 2018.}}
\end{table}	

%%%%%%%%%%%%%%%%%%%%%%%%%%%%%%%%%%%%%%%%%%%

\begin{table}[!ht]\scriptsize \caption[Out-of-sample economic performance of bond excess return forecasts against the EH]{Out-of-sample economic performance of bond excess return forecasts against the EH, measured via certainty equivalent returns (\%) at 1-month prediction horizon - period: January 1985 - end of 2018.}
\label{table:CER}
\begin{center}
\begin{tabular}{lllllll}%{ccccccc}
\toprule % \bf h=1m \backslash
\multicolumn{1}{l}{$\bf n$} &\multicolumn{1}{m{0.8cm}}{\bf 2Y} &\multicolumn{1}{m{0.8cm}}{\bf 3Y} & \multicolumn{1}{m{0.8cm}}{\bf 4Y} & \multicolumn{1}{m{0.8cm}}{\bf 5Y} & \multicolumn{1}{m{0.8cm}}{\bf 7Y} & \multicolumn{ 1}{m{0.8cm}}{\bf 10Y} \\
\midrule                                              
$M_0$\hspace{0.5cm} &     -10.77 &     -11.81 &     -10.27 &      -8.05 &      -3.62 &      -7.49 \\

$M_1$\hspace{0.5cm} &       2.31 &       1.88 &       1.38 &       0.80 &       2.30 &       2.55 \\
\midrule  
\multicolumn{ 7}{c}{\textbf{$CPI$}} \\
\midrule
$LM_{111}$\hspace{0.25cm} &       1.68 &       1.22 &       1.57 &       1.36 &     3.61** &       1.45 \\
$GP_{111}$\hspace{0.25cm} &      3.00* &      2.79* &       2.73 &       2.56 &      4.30* &      3.97* \\
\midrule
$LM_{011}$\hspace{0.25cm} &      2.79* &      2.54* &       2.31 &       1.70 &       3.59 &       2.39 \\
$GP_{011}$\hspace{0.25cm} &     3.69** &     3.30** &      2.91* &       2.50 &      4.01* &      3.69* \\
\midrule
$LM_{110}$\hspace{0.25cm} &     0.74 & 0.02 & 0.07 & 0.12 & 2.56* & 0.06 \\
$GP_{110}$\hspace{0.25cm} &     4.10** & 3.79** & 3.27 & 2.68 & 3.98 & 2.68 \\
% \midrule
% $LM_{101}$\hspace{0.25cm} &   1.78* & 1.49 & 1.97* & 2.03 & 3.85** & 2.91* \\
% $GP_{101}$\hspace{0.25cm} &   3.72** & 3.85** & 3.70* & 3.40 & 4.89* & 5.12** \\
% \midrule
% $LM_{100}$\hspace{0.25cm} &    1.31 & 0.83 & 1.14 & 1.14 & 3.33* & 1.87 \\
% $GP_{100}$\hspace{0.25cm} &    3.27* & 3.07* & 3.00 & 2.54 & 3.71 & 3.45* \\
\midrule
$LM_{010}$\hspace{0.25cm} &    2.22* & 2.07 & 2.13 & 2.09 & 4.33** & 2.28 \\
$GP_{010}$\hspace{0.25cm} &    4.39*** & 4.29** & 4.09** & 3.67* & 4.94* & 3.69** \\
\midrule
$LM_{001}$\hspace{0.25cm} &      2.40* &       2.29 &       2.01 &       1.71 &       3.25 &       2.67 \\
$GP_{001}$\hspace{0.25cm} &     4.14** &    4.43*** &     3.99** &      3.60* &       4.10 &      3.33* \\
\midrule
\multicolumn{ 7}{c}{\textbf{$GRO$}} \\
\midrule
$LM_{111}$\hspace{0.25cm} &      -7.42 &      -1.44 &      -5.44 &      -5.89 &     -10.11 &     -10.29 \\
$GP_{111}$\hspace{0.25cm} &       2.28 &       3.34 &       2.85 &       1.96 &       1.88 &      -0.52 \\
\midrule
$LM_{011}$\hspace{0.25cm} &       3.33 &     3.93** &      3.48* &       3.16 &      5.65* &      -3.89 \\
$GP_{011}$\hspace{0.25cm} &       3.51 &      3.20* &       2.80 &       1.80 &       2.59 &      -2.30 \\
\midrule
$LM_{110}$\hspace{0.25cm} &     -12.12 & -4.54 & -9.08 & -8.39 & -13.46 & -12.49 \\
$GP_{110}$\hspace{0.25cm} &      0.33 & 1.44 & 1.23 & 0.60 & 0.84 & -1.46 \\
% \midrule
% $LM_{101}$\hspace{0.25cm} &     -12.75 & -5.48 & -13.59 & -13.87 & -45.96 & -86.28 \\
% $GP_{101}$\hspace{0.25cm} &     3.65 & 2.95 & 0.70 & -1.39 & -1.65 & 1.35 \\
% \midrule
% $LM_{100}$\hspace{0.25cm} &     -11.67 & -7.80 & -15.19 & -14.71 & -33.00 & -61.51 \\
% $GP_{100}$\hspace{0.25cm} &     2.60 & 2.16 & 0.50 & -1.20 & -0.91 & 2.28 \\
\midrule
$LM_{010}$\hspace{0.25cm} &    2.59 & 3.41 & 2.62 & 2.12 & 5.28* & -2.63 \\
$GP_{010}$\hspace{0.25cm} &    3.45 & 3.11* & 2.50* & 1.96 & 2.73 & -2.86 \\
\midrule
$LM_{001}$\hspace{0.25cm} &     3.61** &     3.37** &     3.75** &       3.50 &       4.29 &       1.60 \\
$GP_{001}$\hspace{0.25cm} &     3.86** &     3.91** &      3.84* &       3.17 &      4.55* &     4.24** \\
\bottomrule
\end{tabular}
\end{center}
\noindent{\caption*{\tiny This table reports annualized certainty equivalent returns ($CER$s) across alternative models at $h$ = 1-month prediction horizon. The coefficient of risk aversion is $\gamma=3$. No portfolio constraints are imposed. $CER$s are generated by out-of-sample forecasts of bond excess returns and are reported in \%. At every time step $t$, an investor with power utility preferences evaluates the entire predictive density of bond excess returns and solves the asset allocation problem, thus optimally allocating her wealth between a riskless bond and risky bonds with maturities 2, 3, 4, 5, 7 and 10-years. $CER$ is then defined as the value that equates the average utility of each alternative model against the average utility of the EH benchmark. The forecasting models used are DTSM with either alternative risk price restrictions or different number of $GP$ outputs with various macros or just other macros. Positive values indicate that the models perform better than the EH benchmark. Statistical significance is measured using a one-sided Diebold-Mariano statistic computed with Newey-West standard errors. * denotes significance at 10\%, ** significance at 5\% and *** significance at 1\% level. The in-sample period is January 1985 to end of 2007, and the out-of-sample period starts in January 2008 and ends in end of 2018.}}
\end{table}

\begin{table}[!ht]\scriptsize \caption[Out-of-sample economic performance of bond excess return forecasts against the corresponding linear model]{Out-of-sample economic performance of bond excess return forecasts against the corresponding linear model, measured via certainty equivalent returns (\%) at 1-month prediction horizon - period: January 1985 - end of 2018.}
\label{table:CERLM}
\begin{center}
\begin{tabular}{lllllll}%{ccccccc}
\toprule % \bf h=1m \backslash
\multicolumn{1}{l}{$\bf n$} &\multicolumn{1}{m{0.8cm}}{\bf 2Y} &\multicolumn{1}{m{0.8cm}}{\bf 3Y} & \multicolumn{1}{m{0.8cm}}{\bf 4Y} & \multicolumn{1}{m{0.8cm}}{\bf 5Y} & \multicolumn{1}{m{0.8cm}}{\bf 7Y} & \multicolumn{ 1}{m{0.8cm}}{\bf 10Y} \\
\midrule  
\multicolumn{ 7}{c}{\textbf{$CPI$}} \\
\midrule
$GP_{111}$\hspace{0.5cm}  &       1.32 &       1.56 &       1.16 &       1.20 &       0.69 &       2.52 \\
$GP_{011}$\hspace{0.5cm}  &      0.89* &       0.76 &       0.60 &       0.79 &       0.42 &       1.30 \\
$GP_{110}$\hspace{0.5cm}  &    3.35*** & 3.78*** & 3.19** & 2.55* & 1.41 & 2.62* \\
% $GP_{101}$\hspace{0.5cm}  &   1.93** & 2.35** & 1.72 & 1.37 & 1.04 & 2.20* \\
% $GP_{100}$\hspace{0.5cm}  &   1.96** & 2.24** & 1.86 & 1.40 & 0.39 & 1.58 \\
$GP_{010}$\hspace{0.5cm}  &   2.16** & 2.21*** & 1.95** & 1.58 & 0.61 & 1.41 \\
$GP_{001}$\hspace{0.5cm}  &     1.88** &     2.54** &     2.34** &     2.17** &       0.61 &      -0.36 \\
\midrule
\multicolumn{ 7}{c}{\textbf{$GRO$}} \\
\midrule
$GP_{111}$\hspace{0.5cm}  &       9.76 &       4.78 &       8.32 &       7.89 &     12.09* &       9.85 \\
$GP_{011}$\hspace{0.5cm}  &       0.17 &      -0.73 &      -0.68 &      -1.35 &      -3.05 &       1.59 \\
$GP_{110}$\hspace{0.5cm}  &      12.58 & 6.00 & 10.39 & 9.06 & 14.46* & 11.15 \\
% $GP_{101}$\hspace{0.5cm}  &     16.57 & 8.46 & 14.45 & 12.63 & 46.08* & 94.42 \\
% $GP_{100}$\hspace{0.5cm}  &     14.40 & 10.03 & 15.89 & 13.68 & 33.00* & 67.23 \\
$GP_{010}$\hspace{0.5cm}  &     0.86 & -0.30 & -0.12 & -0.17 & -2.54 & -0.23 \\
$GP_{001}$\hspace{0.5cm}  &       0.24 &       0.54 &       0.09 &      -0.33 &       0.25 &       2.63 \\
% \midrule
% \multicolumn{ 7}{c}{\textbf{$F8$}} \\
% \midrule
% $GP_{111}$\hspace{0.5cm}  &      1.33* &       0.90 &       1.01 &       1.32 &       0.96 &      -0.54 \\
% $GP_{011}$\hspace{0.5cm}  &       0.11 &      -0.42 &      -0.80 &      -1.02 &      -1.64 &      -1.93 \\
% $GP_{001}$\hspace{0.5cm}  &       0.42 &       0.25 &       0.19 &       0.03 &      -0.92 &       0.09 \\
\bottomrule
\end{tabular}
\end{center}
\noindent{\caption*{\tiny This table reports annualized certainty equivalent returns ($CER$s) across alternative models at $h$ = 1-month prediction horizon. The coefficient of risk aversion is $\gamma=3$. No portfolio constraints are imposed. $CER$s are generated by out-of-sample forecasts of bond excess returns and are reported in \%. At every time step $t$, an investor with power utility preferences evaluates the entire predictive density of bond excess returns and solves the asset allocation problem, thus optimally allocating her wealth between a riskless bond and risky bonds with maturities 2, 3, 4, 5, 7 and 10-years. $CER$ is then defined as the value that equates the average utility of each alternative model against the average utility of the corresponding linear model. The forecasting models used are DTSM with different number of $GP$ outputs and various macros or just other macros. Positive values indicate that the models perform better than the benchmark model. Statistical significance is measured using a one-sided Diebold-Mariano statistic computed with Newey-West standard errors. * denotes significance at 10\%, ** significance at 5\% and *** significance at 1\% level. The in-sample period is January 1985 to end of 2007, and the out-of-sample period starts in January 2008 and ends in end of 2018.}}
\end{table}

\begin{table}[!ht]\scriptsize \caption[Explanatory power of macroeconomic variables when fitting the corresponding nonlinear macros and their components from models $GP_{001}$, $GP_{010}$ and $GP_{110}$]{Explanatory power of macroeconomic variables when fitting the corresponding nonlinear macros and their components from models $GP_{001}$, $GP_{010}$ and $GP_{110}$, measured via $\bar{R}^2$ - period: January 1985 - end of 2017.}
\label{table:R2adjRPFEFPFM001010110}
\begin{center}
\begin{tabular}{c|p{1.3cm}|p{1.3cm}}%|p{0.525cm}p{0.525cm}p{0.525cm}}
\toprule
\multicolumn{3}{c}{$\bar{R}^2:C_{j,t}=a_j+b_jM_{t-1}+e_{j,t},\;j\in\{1,2,3\}$}\\
\midrule
\multicolumn{1}{c|}{$GP_{001}$} & \multicolumn{1}{p{1.3cm}|}{$CPI$} & \multicolumn{1}{p{1.3cm}}{$GRO$} \\
\midrule 
$v^{(3)}(M_{t-1})$ & 0.06 & 0.39 \\%& & 0.01 &\\
& & \\%& & &\\
$E[v^{(3)}(M_{t-1})|\mathcal{P}_t]$ &  0.70 & 0.11 \\%& &     0.00 &\\
$\widetilde{RP}^V_{3,t}(M_{t-1})$  &  0.00 & 0.33 \\%&   &  0.01 &\\
\midrule
\multicolumn{1}{c|}{$GP_{010}$} & \multicolumn{1}{p{1.3cm}|}{$CPI$} & \multicolumn{1}{p{1.3cm}}{$GRO$} \\
\midrule 
$v^{(2)}(M_{t-1})$ &  0.23 &       0.53   \\%& & 0.01 &\\
& & \\%& & &\\
$E[v^{(2)}(M_{t-1})|\mathcal{P}_t]$ &   0.66 &       0.18   \\%& &     0.00 &\\
$\widetilde{RP}^V_{2,t}(M_{t-1})$  &   0.00 &       0.39   \\%&   &  0.01 &\\
\midrule 
\multicolumn{1}{c|}{$GP_{110}$} & \multicolumn{1}{p{1.3cm}|}{$CPI$} & \multicolumn{1}{p{1.3cm}}{$GRO$} \\
\midrule
$v^{(1)}(M_{t-1})$  &     0.73 &       0.25   \\%& &    0.00 &\\
& & \\%& & &\\
$E[v^{(1)}(M_{t-1})|\mathcal{P}_t]$  &   0.70 &       0.17   \\%&  &   0.00 &\\
$\widetilde{RP}^V_{1,t}(M_{t-1})$  &  0.09  &  0.17  \\%&  & 0.00 &\\
\midrule
$v^{(2)}(M_{t-1})$ &  0.32 &  0.45  \\%&  &   0.01 & \\
& & \\%& & &\\
$E[v^{(2)}(M_{t-1})|\mathcal{P}_t]$  &   0.67 &	0.17  \\%& &     0.00 & \\
$\widetilde{RP}^V_{2,t}(M_{t-1})$  &  0.00  &   0.34  \\%& &    0.01 &\\
% \midrule 
% \multicolumn{1}{c|}{$GP_{011}(M)$} & \multicolumn{1}{p{1.3cm}|}{$CPI$} & \multicolumn{1}{p{1.3cm}}{$GRO$} \\
% \midrule
% $f^{(2)}(M_{t-1})$  &    0.44   &  0.50 \\%& &    0.00 &\\
% & & \\%& & &\\
% $E[f^{(2)}(M_{t-1})|\mathcal{P}_t]$  &  0.68  &  0.17 \\%&  &   0.00 &\\
% $\widetilde{RP}^F_{2,t}(M_{t-1})$  & 0.01 & 0.37 \\%&  & 0.00 &\\
% \midrule
% $f^{(3)}(M_{t-1})$ & 0.60 & 0.00 \\%&  &   0.01 & \\
% & & \\%& & &\\
% $E[f^{(3)}(M_{t-1})|\mathcal{P}_t]$  &   0.66  &  0.15 \\%& &     0.00 & \\
% $\widetilde{RP}^F_{3,t}(M_{t-1})$  & 0.07 &  0.03 \\%& &    0.01 &\\
% \midrule 
% \multicolumn{1}{c|}{$GP_{111}(M)$} & \multicolumn{1}{p{1.3cm}|}{$CPI$} & \multicolumn{1}{p{1.3cm}}{$GRO$} \\
% \midrule
% $f^{(1)}(M_{t-1})$  &  0.75  &  0.25  \\%&  &  0.00 &\\
% & & \\%& & &\\
% $E[f^{(1)}(M_{t-1})|\mathcal{P}_t]$  &  0.70 &  0.16  \\%&    & 0.01 &\\
% $\widetilde{RP}^F_{1,t}(M_{t-1})$  &  0.09  & 0.17 \\%& &    0.00 &\\
% \midrule
% $f^{(2)}(M_{t-1})$   & 0.50  &  0.44   \\%&   & 0.01 &\\
% & & \\%& & &\\
% $E[f^{(2)}(M_{t-1})|\mathcal{P}_t]$  & 0.69  & 0.17 \\%&  &   0.00 &\\
% $\widetilde{RP}^F_{2,t}(M_{t-1})$  & 0.02 &  0.32  \\%&    &  0.01 &\\
% \midrule
% $f^{(3)}(M_{t-1})$  &  0.63  &  0.00 \\%&   &   0.01 &\\
% & & \\%& & &\\
% $E[f^{(3)}(M_{t-1})|\mathcal{P}_t]$   & 0.65  & 0.14 \\%&  &   0.00 &\\
% $\widetilde{RP}^F_{3,t}(M_{t-1})$  & 0.08  &  0.04 \\%&  &    0.01 &\\
\bottomrule
\end{tabular}
\end{center}
\noindent{\caption*{\tiny This table reports in-sample $\bar{R}^2$ across alternative regression specifications. The explained variables $C_{j,t}$ are individual nonlinear macros $v^{(j)}(M_{t-1})$ and their components $E[v^{(j)}(M_{t-1})|\mathcal{P}_t]$ and $\widetilde{RP}^V_{j,t}(M_{t-1})$, $j\in\{1,2,3\}$, see \eqref{vDecomposition}, from models $GP_{001}$, $GP_{010}$ and $GP_{110}$. The explanatory variables are the corresponding lagged macroeconomic variables. The sample period is January 1985 to end of 2017.}}
\end{table}

\begin{table}[!ht]\scriptsize \caption[Explanatory power of macroeconomic variables when fitting the corresponding nonlinear macros and their components from models $GP_{011}$ and $GP_{111}$]{Explanatory power of macroeconomic variables when fitting the corresponding nonlinear macros and their components from models $GP_{011}$ and $GP_{111}$, measured via $\bar{R}^2$ - period: January 1985 - end of 2017.}
\label{table:R2adjRPFEFPFM011111}
\begin{center}
\begin{tabular}{c|p{1.3cm}|p{1.3cm}}%|p{0.525cm}p{0.525cm}p{0.525cm}}
\toprule
\multicolumn{3}{c}{$\bar{R}^2:C_{j,t}=a_j+b_jM_{t-1}+e_{j,t},\;j\in\{1,2,3\}$}\\
% \midrule
% \multicolumn{1}{c|}{$GP_{001}(M)$} & \multicolumn{1}{p{1.3cm}|}{$CPI$} & \multicolumn{1}{p{1.3cm}}{$GRO$} \\
% \midrule 
% $f^{(3)}(M_{t-1})$ & 0.06 & 0.39 \\%& & 0.01 &\\
% & & \\%& & &\\
% $E[f^{(3)}(M_{t-1})|\mathcal{P}_t]$ &  0.70 & 0.11 \\%& &     0.00 &\\
% $\widetilde{RP}^F_{3,t}(M_{t-1})$  &  0.00 & 0.33 \\%&   &  0.01 &\\
% \midrule
% \multicolumn{1}{c|}{$GP_{010}(M)$} & \multicolumn{1}{p{1.3cm}|}{$CPI$} & \multicolumn{1}{p{1.3cm}}{$GRO$} \\
% \midrule 
% $f^{(2)}(M_{t-1})$ &  0.23 &       0.53   \\%& & 0.01 &\\
% & & \\%& & &\\
% $E[f^{(2)}(M_{t-1})|\mathcal{P}_t]$ &   0.66 &       0.18   \\%& &     0.00 &\\
% $\widetilde{RP}^F_{2,t}(M_{t-1})$  &   0.00 &       0.39   \\%&   &  0.01 &\\
% \midrule 
% \multicolumn{1}{c|}{$GP_{110}(M)$} & \multicolumn{1}{p{1.3cm}|}{$CPI$} & \multicolumn{1}{p{1.3cm}}{$GRO$} \\
% \midrule
% $f^{(1)}(M_{t-1})$  &     0.73 &       0.25   \\%& &    0.00 &\\
% & & \\%& & &\\
% $E[f^{(1)}(M_{t-1})|\mathcal{P}_t]$  &   0.70 &       0.17   \\%&  &   0.00 &\\
% $\widetilde{RP}^F_{1,t}(M_{t-1})$  &  0.09  &  0.17  \\%&  & 0.00 &\\
% \midrule
% $f^{(2)}(M_{t-1})$ &  0.32 &  0.45  \\%&  &   0.01 & \\
% & & \\%& & &\\
% $E[f^{(2)}(M_{t-1})|\mathcal{P}_t]$  &   0.67 &	0.17  \\%& &     0.00 & \\
% $\widetilde{RP}^F_{2,t}(M_{t-1})$  &  0.00  &   0.34  \\%& &    0.01 &\\
\midrule 
\multicolumn{1}{c|}{$GP_{011}$} & \multicolumn{1}{p{1.3cm}|}{$CPI$} & \multicolumn{1}{p{1.3cm}}{$GRO$} \\
\midrule
$v^{(2)}(M_{t-1})$  &    0.44   &  0.50 \\%& &    0.00 &\\
& & \\%& & &\\
$E[v^{(2)}(M_{t-1})|\mathcal{P}_t]$  &  0.68  &  0.17 \\%&  &   0.00 &\\
$\widetilde{RP}^V_{2,t}(M_{t-1})$  & 0.01 & 0.37 \\%&  & 0.00 &\\
\midrule
$v^{(3)}(M_{t-1})$ & 0.60 & 0.00 \\%&  &   0.01 & \\
& & \\%& & &\\
$E[v^{(3)}(M_{t-1})|\mathcal{P}_t]$  &   0.66  &  0.15 \\%& &     0.00 & \\
$\widetilde{RP}^V_{3,t}(M_{t-1})$  & 0.07 &  0.03 \\%& &    0.01 &\\
\midrule 
\multicolumn{1}{c|}{$GP_{111}$} & \multicolumn{1}{p{1.3cm}|}{$CPI$} & \multicolumn{1}{p{1.3cm}}{$GRO$} \\
\midrule
$v^{(1)}(M_{t-1})$  &  0.75  &  0.25  \\%&  &  0.00 &\\
& & \\%& & &\\
$E[v^{(1)}(M_{t-1})|\mathcal{P}_t]$  &  0.70 &  0.16  \\%&    & 0.01 &\\
$\widetilde{RP}^V_{1,t}(M_{t-1})$  &  0.09  & 0.17 \\%& &    0.00 &\\
\midrule
$v^{(2)}(M_{t-1})$   & 0.50  &  0.44   \\%&   & 0.01 &\\
& & \\%& & &\\
$E[v^{(2)}(M_{t-1})|\mathcal{P}_t]$  & 0.69  & 0.17 \\%&  &   0.00 &\\
$\widetilde{RP}^V_{2,t}(M_{t-1})$  & 0.02 &  0.32  \\%&    &  0.01 &\\
\midrule
$v^{(3)}(M_{t-1})$  &  0.63  &  0.00 \\%&   &   0.01 &\\
& & \\%& & &\\
$E[v^{(3)}(M_{t-1})|\mathcal{P}_t]$   & 0.65  & 0.14 \\%&  &   0.00 &\\
$\widetilde{RP}^V_{3,t}(M_{t-1})$  & 0.08  &  0.04 \\%&  &    0.01 &\\
\bottomrule
\end{tabular}
\end{center}
\noindent{\caption*{\tiny This table reports in-sample $\bar{R}^2$ across alternative regression specifications. The explained variables $C_{j,t}$ are individual nonlinear macros $v^{(j)}(M_{t-1})$ and their components $E[v^{(j)}(M_{t-1})|\mathcal{P}_t]$ and $\widetilde{RP}^V_{j,t}(M_{t-1})$, $j\in\{1,2,3\}$, see \eqref{vDecomposition}, from models $GP_{011}$ and $GP_{111}$. The explanatory variables are the corresponding lagged macroeconomic variables. The sample period is January 1985 to end of 2017.}}
\end{table}

\end{document}